\documentclass[pra,twocolumn,groupedaddress,showpacs,floatfix]{revtex4}
\pdfoutput=1
\usepackage{latexsym,amsfonts,amssymb,amsmath,graphicx,graphics,hyperref}
\newcommand{\ket}[1]{\left| #1 \right\rangle}
\newcommand{\bra}[1]{\left\langle #1 \right|}

\begin{document}

\title{Enhancing laser sideband cooling in one-dimensional optical lattices \\ via the dipole interaction}
\author{Rebecca N. Palmer}
\email{R.Palmer@leeds.ac.uk}
\author{Almut Beige}
\affiliation{The School of Physics and Astronomy, University of Leeds, Leeds, LS2 9JT, United Kingdom} 

\date{\today}

\begin{abstract}
We study resolved sideband laser cooling of a one-dimensional optical lattice with one atom per site, and in particular the effect of the dipole interaction between radiating atoms. For simplicity, we consider the case where only a single cooling laser is applied.  We derive a master equation, and solve it in the limit of a deep lattice and weak laser driving.  We find that the dipole interaction significantly changes the final temperature of the particles, increasing it for some phonon wavevectors and decreasing it for others. The total phonon energy over all modes is typically higher than in the non-interacting case, but can be made lower by the right choice of parameters.
\end{abstract}

\pacs{37.10.De,37.10.Jk}

\maketitle

\section{Introduction} 
 
Over the recent decades, a variety of laser cooling techniques have been developed \cite{laser-cooling-limits,narrowline-doppler-theory,polarization-gradient-theory,sw-sideband-theory} and implemented \cite{bec-orig-Rb,bec-orig-Na,narrowline-doppler-Sr,raman-doppler-Na,raman-doppler-Cs} for cooling trapped atoms or ions to very low temperatures.  The resulting ultracold atoms can be made into a Bose-Einstein condensate \cite{bec-orig-Rb,bec-orig-Na}.  They can also be put into an optical lattice (an optical standing wave forming a periodic potential \cite{lattice-review-Jaksch05,lattice-review09,lattice-qs-review}), where they can be used for quantum simulation of condensed matter systems \cite{lattice-qs-review}, or potentially for quantum information processing \cite{lattice-review-Jaksch05}.  However, laser cooling does have limitations, which often require it to be followed by the loss of $\sim 99-99.9\%$ of the atoms by evaporative cooling \cite{bec-orig-Rb,bec-orig-Na}.

\begin{figure}[t]
\includegraphics{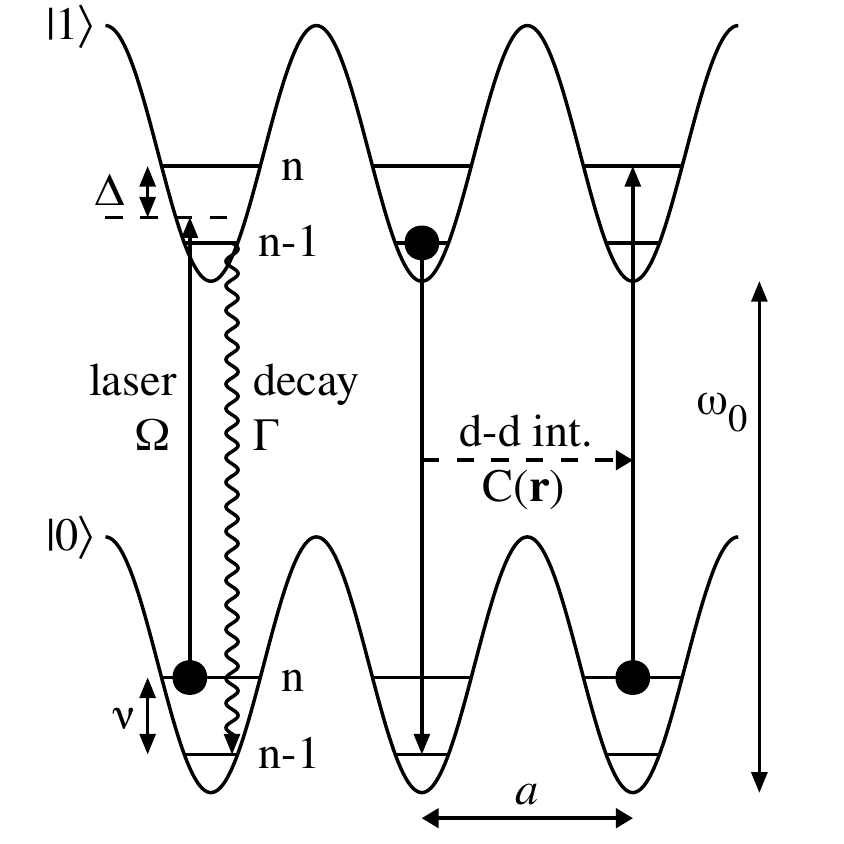}
\vspace*{-0.5cm} \caption{Energy level diagram of our system.  In the left site we show the conventional sideband cooling cycle: for red detuning $\Delta\approx\nu$, the resonant excitation process is to the level with one fewer phonons (motional quanta), while in a deep lattice the dominant spontaneous decay process does not change the phonon number.  On the right, we show an example of dipole-dipole interaction: the photon emitted by the middle atom is reabsorbed by the right-hand one, moving the excitation from the middle to the right-hand site.  In this example the phonon states of both atoms are unchanged, but we will see later that this is not always the case.} \label{dipolecoollevels}
\end{figure}

We here consider resolved sideband cooling \cite{laser-cooling-limits,sw-sideband-theory}, which is mostly used on ions \cite{ion-cooling-review} but has been applied to atoms in an optical lattice \cite{sideband-3dlattice-Cs,sideband-1dlattice-Cs}. This cooling technique works on strongly confined particles: the trapping frequency $\nu$ needs to be large compared to the spontaneous decay rate $\Gamma$ of the internally excited state. This allows the quantized motional states (with spacing $\hbar\nu$) to be individually resolved.  We can hence tune the driving laser to drive mainly a transition to the next lower motional level (Fig.~\ref{dipolecoollevels}), removing one motional quantum per cooling cycle.

Equilibrium is reached when this cooling is balanced by heating.  For a single particle, heating can occur either by off-resonant excitation or by the random recoil of spontaneous emission \cite{laser-cooling-limits}.  As confined particles can only be heated in whole quanta $\hbar\nu$, heating is rare if $\hbar\nu$ is large compared to the widths of these heating processes, $\hbar\Gamma$ and $E_R\equiv\hbar^2 k_0^2/2m$ respectively (here $k_0$ is the transition wavenumber and $m$ is the atomic mass). This gives an equilibrium temperature of $\sim\hbar\Gamma^2/\nu$, compared to $\sim\hbar\Gamma$ for free space Doppler cooling \cite{laser-cooling-limits}.  Resolved sideband cooling also requires only one laser frequency, rather than the spread of frequencies required for narrow-line ($\hbar\Gamma<E_R$) Doppler cooling \cite{narrowline-doppler-theory}.

% A particle emitting or absorbing dipole radiation has an oscillating dipole, and if other particles are present these dipoles interact.  This interaction modifies the decay rate, which is called superradiance \cite{two-dipole1,two-dipole2,Agarwal}.  It also allows reabsorption of photons, which is an additional source of heating \cite{reabsorption-heating}, but like single atom heating is suppressed by strong confinement \cite{reabsorption-lattice,reabsorption-heating}.

When laser cooling is applied to many closely spaced particles, the spontaneous decay rate is modified by interference between the emission from different particles, which is called superradiance \cite{two-dipole1,two-dipole2,Agarwal}.  It is also possible for a photon emitted by one atom to be reabsorbed by another \cite{reabsorption-heating}.  These two effects together are the dipole-dipole interaction we consider.  This interaction exists between any particles emitting electric dipole radiation (the dipole is the matrix element of the transition), and becomes significant when the distance between them is of the order of the transition wavelength.  It modifies the decay rate and propagates electronic excitations from atom to atom, and has no effect when no electronic excitations are present.  It is hence qualitatively distinct from static interparticle forces, such as the Coulomb interaction between ions \cite{ion-chain-cooling} or the static-dipole interaction between aligned polar molecules \cite{dipole-gas-review08}, which propagate motional excitations (phonons).  Previous treatments of this interaction have considered either the superradiance part only \cite{collective-darkstate-cooling}, the reabsorption part only \cite{reabsorption-lattice}, or a long range approximation \cite{reabsorption-heating,rescattering-force-theory} which is valid for a moderate density gas but breaks down at the half-wavelength spacing typical of an optical lattice.

\begin{figure}[t]
\includegraphics{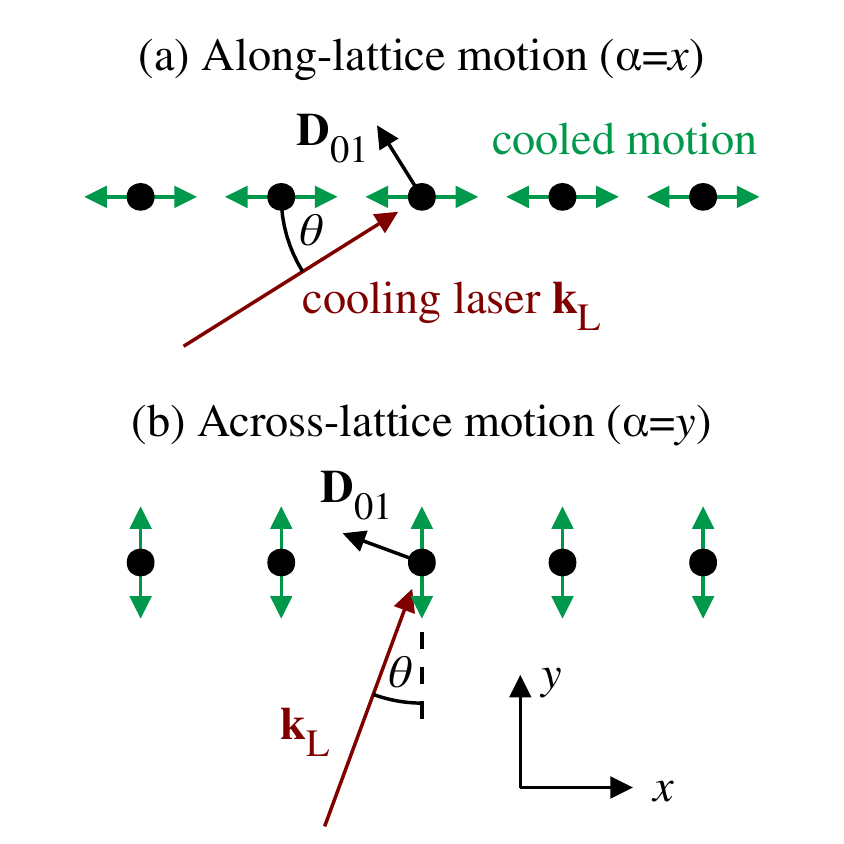}
\vspace*{-0.5cm} \caption{(Color online) Physical layout of our system.  We consider a one-dimensional array of tightly confined atoms, one per lattice site.  For simplicity we assume only one cooling laser, and hence can cool the atomic motion in any one dimension, either along the lattice (which we call $x$) or across it ($y$ or $z$), but not all three at once. As in the single particle case, cooling all three dimensions would require multiple lasers.} \label{dipolecoollayout}
\end{figure}

In this paper we consider a deep one-dimensional optical lattice with one atom per site, as shown in Fig.~\ref{dipolecoollayout}, and include the full dipole-dipole interaction.  We solve for the steady state by transforming to momentum space and using a diagrammatic perturbative expansion in the Lamb-Dicke parameter $\eta\equiv\sqrt{E_R/\nu}$.  We find that the dipole-dipole interaction may increase or decrease the steady state phonon number, depending on the laser parameters.  With optimized settings, the phonon number can be made $\approx 30\%$ lower than would be possible without interaction.

There are six sections in this paper. Section \ref{exp} derives the relevant master equation for two-level atoms tightly confined by an optical lattice, including the dipole-dipole interaction. In Section \ref{fourier} we simplify this equation for the case of an infinite one-dimensional optical lattice by transforming to momentum space.  In Section \ref{stat}, we introduce a Feynman-like diagrammatic perturbative expansion, and use it to solve for the steady state phonon number distribution to lowest order in the Lamb-Dicke parameters $\eta_{\alpha}$. The Appendix does the same expansion using non-diagrammatic methods. In Section \ref{examples}, we evaluate this steady state phonon distribution for particular lattice setups, and find the parameter settings that minimize the phonon number.  Finally, Section \ref{conc} gives conclusions.

\section{Theoretical model} \label{exp}

In this section, we derive the master equation of our system, including both the dipole-dipole interaction between the atoms and the coupling between their electronic and motional states.

Our starting point is the dipole Hamiltonian, which includes energy terms for the electronic and motional (phonon) excitations of the atoms and for free photons, and interaction terms between the atomic dipole moments and the electric fields of the cooling laser and the free photons.  We then eliminate the photons to obtain the spontaneous decay and dipole-dipole interaction, using similar methods to Refs.~\cite{Agarwal,two-dipole1,two-dipole2}.  However, they consider atoms in fixed positions, while we allow the atoms to move within their lattice sites.  We hence obtain additional terms in the dipole-dipole interaction, describing processes that change the motional state as well as the electronic state. We derive explicit expressions for the corresponding coupling constants up to second order in the Lamb-Dicke parameters.

\subsection{The Schr\"odinger picture}

Our system consists of $N$ two-level atoms in a deep 1D optical lattice with one atom per site, as shown in Figs.~\ref{dipolecoollevels},\ref{dipolecoollayout}.  We denote the two internal levels of atom $i$ by $\ket{0}_i$ and $\ket{1}_i$, and the transition energy between them by $\hbar \omega_0$.  We then define the atom raising and lowering operators 
\begin{eqnarray} \label{S's}
S^\dagger_i \equiv \ket{1}_{ii}\bra{0} ~~ \mathrm{and} ~~ S_i \equiv \ket{0}_{ii}\bra{1} \, . 
\end{eqnarray}

We assume that the lattice is sufficiently deep that tunneling between neighboring lattice sites is negligible (i.e. atom $i$ stays in site $i$), and each site can be approximated by a harmonic well.  We denote the trapping frequency of this harmonic well by $\nu_\alpha$, which we assume to be the same for both internal states, but allow to be anisotropic, where $\alpha=x,y,z$ denotes the direction.  However, unlike the usual Bose-Hubbard model \cite{lattice-review-Jaksch05}, we do not assume that the atoms are in the lowest motional band of the lattice, but allow them to be in higher motional states.  The excitations to these higher motional states are the phonons, with creation and annihilation operators $B_{i \alpha}^\dagger,B_{i \alpha}$ for a motional excitation of atom $i$ in the direction $\alpha$.

For simplicity we consider the case of a single plane polarized cooling laser, with peak electric field $\mathbf{E}_\mathrm{L}$, wavevector $\mathbf{k}_\mathrm{L}$, and frequency $\omega_\mathrm{L}$.  To describe the spontaneous decay and dipole-dipole interaction, we also need to include the free radiation field, with photon annihilation operators $a_{\mathbf{k}\lambda}$ labelled by a wavevector $\mathbf{k}$ and a polarization index $\lambda=1,2$.

We treat the atom-field interactions in the electric dipole approximation, i.e. as the dot product of the atomic dipole moment
\begin{eqnarray}
\mathbf{D}_i &=&\mathbf{D}_{01} \, S_i + \mathrm{H.c.}
\end{eqnarray}
and the electric field.  The classical laser field is
\begin{eqnarray}
\mathbf{E}_\mathrm{L}(\mathbf{x},t) &=& \mathbf{E}_\mathrm{L} \cos(\mathbf{k}_\mathrm{L}\cdot \mathbf{x} - \omega_\mathrm{L} t)\, ,
\end{eqnarray}
while the free photons have electric field operator
\begin{eqnarray}
\mathbf{E}(\mathbf{x}) &=& \sum_{\mathbf{k}\lambda} \mathrm{i} \sqrt{{\omega_k \over 2 \epsilon_0 \hbar L^3}} \, \mathbf{\epsilon}_{\mathbf{k}\lambda} \, \mathrm{e}^{\mathrm{i} \mathbf{k} \cdot \mathbf{x}} \, a_{\mathbf{k}\lambda} + \mathrm{H.c.} \, ,
\end{eqnarray}
where $L^3$ is the quantization volume and $\mathbf{\epsilon}_{\mathbf{k}\lambda}$ the polarization vector.

Putting all these together, the Hamiltonian of the system in the Schr\"odinger picture is
\begin{eqnarray}
H&=&\sum\limits_{i=1}^N\hbar\omega_0 \, S^\dagger_iS_i+\sum\limits_{i=1}^N e\mathbf{D}_i\cdot \big[ \mathbf{E}_\mathrm{L}(\mathbf{x}_i,t) + \mathbf{E}(\mathbf{x}_i) \big] \notag \\
&& + \sum_{\mathbf{k}\lambda} \hbar\omega_k \, a_{\mathbf{k}\lambda}^\dagger a_{\mathbf{k}\lambda} +\sum\limits_{i=1}^N \sum_{\alpha=x,y,z} \hbar \nu_\alpha \, B_{i \alpha}^\dagger B_{i \alpha} \, ,
\end{eqnarray}
where $\mathbf{x}_i$ is the position of atom $i$.
\subsection{The interaction picture}

We next eliminate the time dependence of the laser term, by moving to the interaction picture with respect to
\begin{equation}
H_0=\sum\limits_{i=1}^N\hbar\omega_\mathrm{L} \, S^\dagger_i S_i + \sum_{\mathbf{k}\lambda} \hbar\omega_k \, a_{\mathbf{k}\lambda}^\dagger a_{\mathbf{k}\lambda} \, ,
\end{equation}
and applying the rotating wave approximation with respect to $\omega_0$ (i.e. neglecting terms $\mathrm{e}^{\mathrm{i}\omega t}$ with $\omega\gtrsim\omega_0$).  This gives the interaction picture Hamiltonian
\begin{eqnarray}
H_\mathrm{I}&=& \sum_{i=1}^N\hbar \Delta \, S^\dagger_i S_i + \sum\limits_{i=1}^N \left[{1 \over 2} \hbar \Omega \, \mathrm{e}^{- \mathrm{i} \mathbf{k}_\mathrm{L}\cdot \mathbf{x}_i} \, S_i\right.\notag\\
&&\hspace*{-0.5cm}\left.- \sum_{\mathbf{k}\lambda}  \mathrm{i} e \sqrt{{\omega_k \over 2 \epsilon_0 \hbar L^3}} \, \mathbf{D}_{01} \cdot \mathbf{\epsilon}_{\mathbf{k}\lambda}^* \, \mathrm{e}^{- \mathrm{i} \mathbf{k} \cdot \mathbf{x}_i} \, 
\mathrm{e}^{\mathrm{i}(\omega_k - \omega_\mathrm{L})t}S_i \, a_{\mathbf{k}\lambda}^\dagger \right]\notag \\
&&\hspace*{-0.5cm}+\, \mathrm{H.c.}\, + \sum_{i=1}^N \sum_{\alpha=x,y,z} \hbar \nu_\alpha \, B_{i \alpha}^\dagger B_{i \alpha}\, ,
\end{eqnarray}
where we define the laser detuning $\Delta \equiv \omega_0 - \omega_\mathrm{L}$ (i.e. red detuning is positive) and the Rabi frequency
\begin{eqnarray}
\Omega &\equiv & {2 e \over \hbar} \, \mathbf{D}_{01} \cdot \mathbf{E}_\mathrm{L}^* \, .
\end{eqnarray}
We choose the relative phase of $|0 \rangle$, $|1 \rangle$ so as to make $\Omega$ real. 

\subsection{The master equation}

As our system is not in an optical cavity, the free photons $a_{\mathbf{k}\lambda}$ quickly leave the system, and are absorbed by the walls of the apparatus on a time scale $\Delta t\sim L/c$, making our system an open quantum system.  We hence need to use a master equation rather than a Hamiltonian to describe its evolution.  For realistic sizes, $\omega_0\gg 1/\Delta t\gg \nu,\Delta,\Omega,\Gamma$ (where $\Gamma$ is the decay rate derived below), which allows us to use the method of Refs.~\cite{reset,two-dipole2,two-dipole1} to derive a Markovian (memoryless) master equation.

This method approximates the photon loss process by alternating free evolution under $H_\mathrm{I}$ for time $\Delta t$ with the instantaneous loss of all photons,
\begin{eqnarray}
\rho(t_0+\Delta t)&=&\mathrm{Tr}_\mathrm{photons}\left[U_\mathrm{I}(t_0+\Delta t,t_0)\right.\notag\\
&&\hspace*{-1cm}\left.\times\left\{\rho(t_0)\otimes\ket{0}\bra{0}_\mathrm{pht}\right\} U_\mathrm{I}^\dagger(t_0+\Delta t,t_0)\right],
\end{eqnarray}
where $\rho(t)$ is the (electronic and motional) density matrix of the atoms at time $t$, $\ket{0}_\mathrm{pht}$ is the vacuum state of the photon field, and $U_\mathrm{I}(t_0+\Delta t,t_0)$ is the unitary evolution operator under $H_\mathrm{I}$ from time $t_0$ to $t_0+\Delta t$.

Since $\Delta t$ is short compared to the dynamical timescales of $H_\mathrm{I}$, we can approximate this by expanding the time ordered exponential $U_\mathrm{I}(t_0+\Delta t,t_0)=\mathcal{T}\exp\left\{-\frac{\mathrm{i}}{\hbar}\int\limits_{t_0}^{t_0+\Delta t} dt H_\mathrm{I}(t)\right\}$ up to second order in $\Delta t$:

\begin{eqnarray}\label{master-exp}
\frac{\Delta\rho}{\Delta t}&=&\mathrm{Tr}_\mathrm{photons}\left[-\frac{\mathrm{i}}{\hbar}\left\{\frac{1}{\Delta t}\int\limits_{t_0}^{t_0+\Delta t} dt H_\mathrm{I}(t)\right.\right.\\
&&\hspace*{-1.5cm}\left.-\frac{\mathrm{i}}{\hbar\Delta t}\int\limits_{t_0}^{t_0+\Delta t} dt\int\limits_{t_0}^{t} dt^\prime H_\mathrm{I}(t)H_\mathrm{I}(t^\prime)\right\}\rho(t_0)\otimes\ket{0}\bra{0}_\mathrm{pht}+\mathrm{H.c.}\notag\\
&&\hspace*{-1.5cm}\left.+\frac{1}{\hbar^2\Delta t}\int\limits_{t_0}^{t_0+\Delta t} dt\int\limits_{t_0}^{t_0+\Delta t} dt^\prime H_\mathrm{I}(t)\rho(t_0)\otimes\ket{0}\bra{0}_\mathrm{pht}H_\mathrm{I}(t^\prime)\right]\, ,\notag
\end{eqnarray}
where $\Delta\rho=\rho(t_0+\Delta t)-\rho(t_0)$.  In the limit $\omega_0\gg 1/\Delta t\gg \nu,\Delta,\Omega,\Gamma$, the above integrals can be evaluated and give a time-independent master equation, which can be conveniently written as \cite{two-dipole1,two-dipole2}
\begin{eqnarray} \label{master}
\dot {\rho} = - {\mathrm{i} \over \hbar} \big[ \, H_\mathrm{cond} \, \rho - \rho \, H_\mathrm{cond}^\dagger \, \big] + \mathcal{R} (\rho) \, ,
\end{eqnarray}
where $H_\mathrm{cond}$ comes from the terms of (\ref{master-exp}) in curly brackets and $\mathcal{R} (\rho)$ comes from the last term.
\subsection{The no-photon time evolution}

The term $H_\mathrm{cond}$ in Eq.~(\ref{master}) describes processes where no photons are lost from the system, either because none are emitted, or because one is emitted but is reabsorbed before time $\Delta t$ \cite{two-dipole1,two-dipole2}.  As it appears in the master equation (\ref{master}) in a similar form to the $-\frac{\mathrm{i}}{\hbar}[H,\rho]$ of a Hamiltonian, but describes only evolution under the condition of no photon loss, this term is called the conditional Hamiltonian.  Unlike a normal Hamiltonian, it is not Hermitian, but norm decreasing; this represents the decreasing probability that this condition will continue to hold, and in the master equation is balanced by $\mathcal{R} (\rho)$.

Evaluating (\ref{master-exp}) using the methods of \cite{two-dipole1,two-dipole2}, we find
\begin{eqnarray} \label{Hcond}
H_\mathrm{cond} &=& \sum\limits_{i=1}^N \Bigg[ {1 \over 2} \hbar \Omega \, \mathrm{e}^{- \mathrm{i} \mathbf{k}_\mathrm{L}\cdot \mathbf{x}_i} \, S_i + \mathrm{H.c.} \notag \\
&& + \hbar \Big( \Delta - {\mathrm{i} \over 2} \Gamma \Big) \, S^\dagger_i S_i + \sum_{\alpha=x,y,z} \hbar \nu_\alpha \, B_{i \alpha}^\dagger B_{i \alpha} ~~~ \notag \\
&& - \sum_{j \neq i} \frac{\mathrm{i}}{2} \hbar C(\mathbf{x}_j - \mathbf{x}_i ) \, S^\dagger_j \, S_i \Bigg] \, .
\end{eqnarray}
The terms of (\ref{Hcond}) describe respectively laser driving, detuning and spontaneous decay, phonon energy, and dipole-dipole interaction.  All except the last are identical to standard laser cooling.

The last term contains the dipole coupling operator
\begin{eqnarray} \label{Cij}
C(\mathbf{r}) &=& {3 \over 2} \, \Gamma \, \mathrm{e}^{\mathrm{i} k_0 r} \, \Bigg[ \frac{1}{\mathrm{i} k_0 r} \big(1 -| \mathbf{\hat D}_{01} \cdot \mathbf{\hat r} |^2 \big) \notag \\
&&\hspace*{-0.5cm} +\left(\frac{1}{(k_0 r)^2}-\frac{1}{\mathrm{i} (k_0 r)^3} \right) \big( 1 -3 \, |\mathbf{\hat D}_{01} \cdot \mathbf{\hat r} |^2 \big) \Bigg] \, ,
\end{eqnarray}
where $r \equiv \| \mathbf{r} \|$, $\mathbf{\hat r} \equiv \mathbf{r}/ r$, $k_0 \equiv \omega_0/c$ and $\mathbf{\hat D}_{01} \equiv \mathbf{D}_{01}/ \| \mathbf{D}_{01} \|$.  The imaginary part of $C(\mathbf{x}_j - \mathbf{x}_i )$ describes photons being emitted by atom $i$ and reabsorbed by atom $j$, while the real part describes modifications to the decay rate due to interference between their emitted waves.  Eq.~(\ref{Cij}) is exactly the same expression as the dipole coupling constants in Refs.~\cite{Agarwal,two-dipole1,two-dipole2}, but they fix the atom positions so their $\mathbf{r}$ is a classical vector, while we allow motion within each lattice site and our $\mathbf{r}$ is hence an operator.

\subsection{Lamb-Dicke expansion}

Eq.~(\ref{Hcond}) is difficult to use as it stands because it contains transcendental functions of the operator $\mathbf{x}_i$, namely $\exp(- \mathrm{i} \mathbf{k}_\mathrm{L}\cdot \mathbf{x}_i)$ and $C(\mathbf{x}_j-\mathbf{x}_i)$.  In this section we introduce the Lamb-Dicke expansion about the site centers, which allows these to be handled perturbatively in the deep lattice limit.

\begin{figure}[t]
\includegraphics{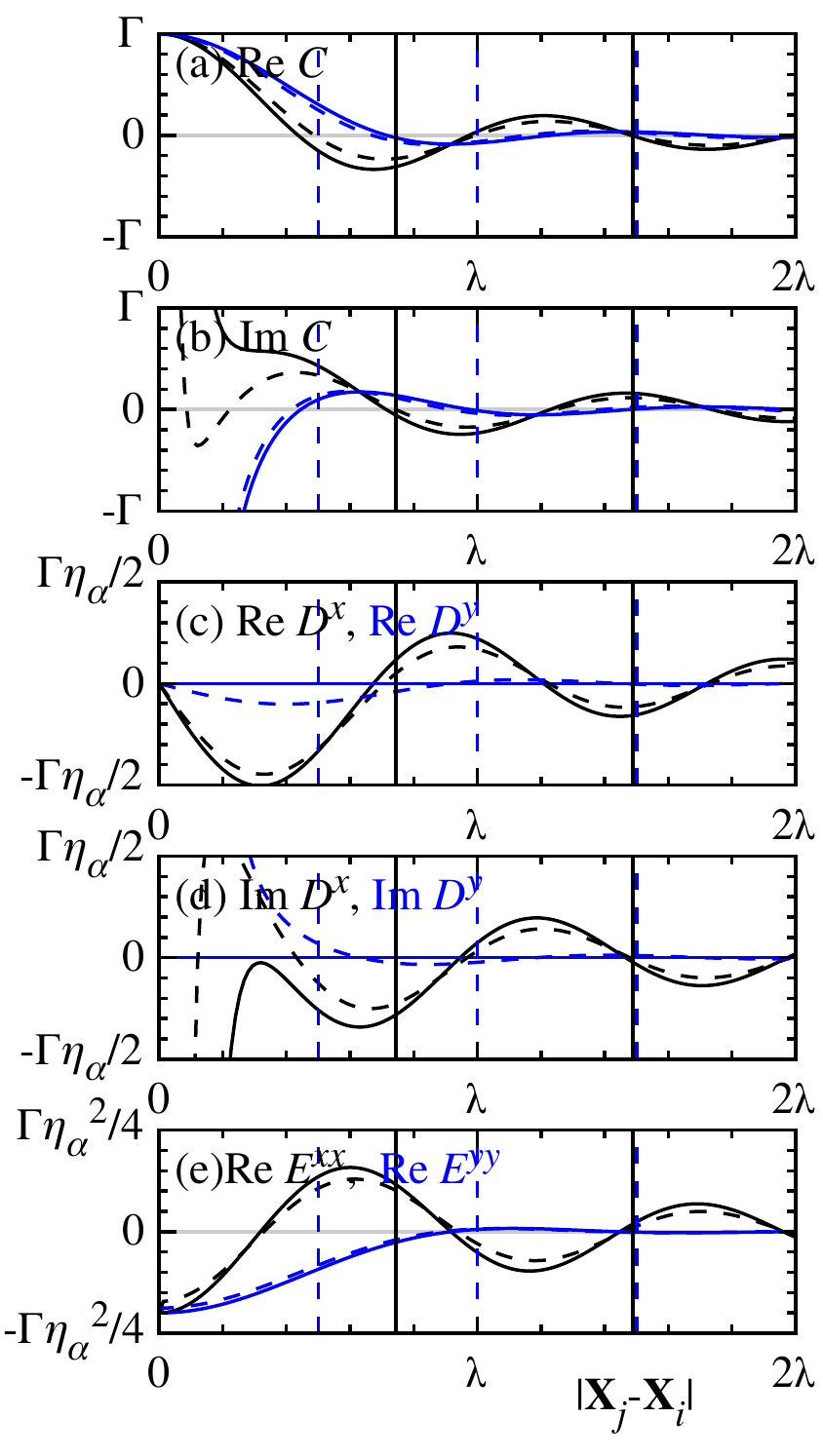}
\vspace*{-0.5cm} \caption{(Color online) Dipole coupling constants $C(\mathbf{X})$, $D^\alpha (\mathbf{X})$, and $E^{\alpha \alpha}(\mathbf{X})$, as a function of distance $|\mathbf{X}|$, for along-lattice motion (black) and across-lattice motion (blue).  Solid lines are $\vartheta=0$ ($\mathbf{\hat D}_{01} \cdot \mathbf{\hat X}=\sin\vartheta$ for along-lattice motion or $\cos\vartheta$ for across-lattice motion, cf.~Fig.~\ref{dipolecoollayout}).  Dotted lines are $\vartheta=0.56$ (along-lattice) or $0.35$ (across-lattice), which we will see later are the optimum values at lattice spacing $a=0.5\lambda$.  For $D$ and $E$ the component plotted is the relevant one for cooling that motion direction: $\alpha=x$ for along-lattice motion or $\alpha=y$ for across-lattice motion.  The vertical lines mark the lattice positions for $a=0.5\lambda$ (blue dotted) and $a=0.744\lambda$ (black solid).} \label{Cplot}
\end{figure}

As already mentioned above, we approximate each lattice site by a harmonic well with phonon operators $B_{i \alpha}, B_{i \alpha}^\dagger$.  In terms of these operators, the position operator $\mathbf{x}_i$ is 
\begin{eqnarray} \label{xi}
{x}_{i\alpha} &= & {X}_{i\alpha} + \frac{\eta_\alpha}{k_0} \big(B_{i \alpha} + B_{i \alpha}^\dagger \big)\, , ~~~
\end{eqnarray}
where $\mathbf{X}_i$ is the equilibrium position (site center), $\alpha=x,y,z$ labels vector components, and the Lamb-Dicke parameter $\eta_\alpha$ is defined by 
\begin{eqnarray} 
\eta_\alpha &\equiv & \sqrt{\hbar k_0^2 / 2 m \nu_\alpha} \, ,
\end{eqnarray}
where $m$ is the mass of the atom.  In the case of a deep lattice $\eta_\alpha$ is small, and we can hence use Eq.~(\ref{xi}) to approximate $\exp(- \mathrm{i} \mathbf{k}_\mathrm{L}\cdot \mathbf{x}_i)$ and $C(\mathbf{x}_j-\mathbf{x}_i)$ in Eq.~(\ref{Hcond}) by Taylor expansion.  Up to second order in $\eta_{\alpha}$, we find 
\begin{eqnarray} \label{expansion}
\mathrm{e}^{- \mathrm{i} \mathbf{k}_\mathrm{L} \cdot \mathbf{x}_i} &=& \mathrm{e}^{- \mathrm{i} \mathbf{k}_\mathrm{L}\cdot \mathbf{X}_i} \, \Bigg[ 1 - \mathrm{i} \sum_{\alpha=x,y,z} \hat k_{\mathrm{L}\alpha}\eta_\alpha \big(B_{i \alpha} + B_{i \alpha}^\dagger \big) \notag \\
&& \hspace*{-1.6cm} - \hspace*{-0.2cm}\sum_{\alpha, \beta=x,y,z}\hspace*{-0.2cm}\frac{\eta_\alpha \eta_\beta}{2} \hat k_{\mathrm{L}\alpha}\hat k_{\mathrm{L}\beta}\big(B_{i \alpha} + B_{i \alpha}^\dagger \big) \big(B_{i \beta} + B_{i \beta}^\dagger \big) \Bigg] \, , ~~~
\end{eqnarray}
where $\mathbf{\hat k}_\mathrm{L}\equiv\mathbf{k}_\mathrm{L}/|\mathbf{k}_\mathrm{L}|$ and $\hat k_{\mathrm{L}\alpha}$ is the component of this unit vector in the $\alpha$ direction, and
\begin{eqnarray} \label{constants}
C(\mathbf{x}_j-\mathbf{x}_i) &=& C_{ij} \notag \\
&& \hspace*{-1.5cm} + \sum_{\alpha=x,y,z} D^\mathrm{\alpha}_{ij} \big( B_{j \alpha} + B_{j \alpha}^\dagger - B_{i \alpha} - B_{i \alpha}^\dagger \big) \notag \\
&& \hspace*{-2cm} + \hspace*{-0.2cm}\sum_{\alpha,\beta=x,y,z}\hspace*{-0.2cm} E^{\alpha \beta}_{ij} \big( B_{j \alpha} + B_{j \alpha}^\dagger - B_{i \alpha} - B_{i \alpha}^\dagger \big) \notag \\
&& \hspace*{-1.5cm} \times \big( B_{j \beta} + B_{j \beta}^\dagger - B_{i \beta} - B_{i \beta}^\dagger \big). 
\end{eqnarray}
Here the dipole coupling constants $C_{ij}$, $D^\alpha_{ij}$, and $E^{\alpha\beta}_{ij}$ are obtained by differentiating $C(\mathbf{r})$ (and absorbing the $\eta_\alpha/k_0$ factors),
\begin{eqnarray} \label{Dij}
D^\mathrm{\alpha}(\mathbf{r}) &\equiv & {\eta_\alpha \over k_0} \cdot {\partial C(\mathbf{r}) \over {\partial r_\alpha}} \, , \notag \\
E^\mathrm{\alpha \beta}(\mathbf{r}) &\equiv & {\eta_\alpha \eta_\beta \over 2 k_0^2} \cdot {\partial^2 C(\mathbf{r}) \over {\partial r_\alpha\partial r_\beta}} \, ,
\end{eqnarray}
where $r_\alpha$ are the components of $\mathbf{r}$, then substituting the equilibrium positions $\mathbf{X}_i$ for the atoms,
\begin{eqnarray} \label{Cij2}
C_{ij} &\equiv & C(\mathbf{X}_j-\mathbf{X}_i) \, ,\nonumber\\
D^\alpha_{ij}& \equiv &D^\alpha(\mathbf{X}_j-\mathbf{X}_i)\, ,\nonumber\\
E^{\alpha\beta}_{ij}& \equiv &E^{\alpha\beta}(\mathbf{X}_j-\mathbf{X}_i)\, .
\end{eqnarray}
Because they contain only equilibrium positions $\mathbf{X}_i$ and not position operators $\mathbf{x}_i$, these are numbers rather than operators.  Finally we remark that $C_{ij} = C_{ji}$, and hence $D_{ij}^\alpha = - D_{ji}^\alpha$ and $E_{ij}^{\alpha\beta} = E_{ji}^{\alpha\beta}$.  In Fig.~\ref{Cplot} we plot these coupling constants against the distance $|\mathbf{X}_j-\mathbf{X}_i|$ between the atoms, for a few relevant angles between $\mathbf{X}_j-\mathbf{X}_i$ and the dipole moment $\mathbf{\hat D}_{01}$. 

Substituting Eqs.~(\ref{expansion}) and (\ref{constants}) back into the conditional Hamiltonian (\ref{Hcond}), we see that in zeroth order in $\eta_\alpha$, the dipole-dipole coupling constants $C_{ij}$ are the same as for atoms in fixed positions \cite{Agarwal,two-dipole1,two-dipole2}, while the motional states are unaffected.  The first and higher order terms couple the electronic and motional states of the atoms: when atoms exchange a photon, they may also create or destroy one or more phonons, with coupling constants $D^\mathrm{\alpha}_{ij}$ and $E^\mathrm{\alpha \beta}_{ij}$.

\subsection{The effect of spontaneous emission}

The term $\mathcal{R}(\rho)$ in Eq.~(\ref{master}) describes processes where photon loss does occur; it is called the reset operator as it ``resets" the system state after such a loss \cite{reset}.  To evaluate this expression we proceed as in Refs.~\cite{two-dipole1,two-dipole2}, and again obtain something similar to the fixed-positions expression $\mathrm{Re}\,C(\mathbf{x}_j - \mathbf{x}_i ) \, S_i \, \rho \, S^\dagger_j$ with the positions becoming operators.  However, here $\mathrm{Re}\,C(\mathbf{x}_j - \mathbf{x}_i )$ does not simply act from the left, but becomes a superoperator (i.e. contains operators acting from both sides).  In Lamb-Dicke expansion, each phonon operator acts from the same side of $\rho$ as the same atom's electronic (de)excitation operator, i.e. $B_{i \alpha} + B_{i \alpha}^\dagger$ on the left and $B_{j \alpha} + B_{j \alpha}^\dagger$ on the right.  Up to second order in $\eta_\alpha$, this gives 
\begin{eqnarray} \label{reset}
\mathcal{R}(\rho) &=& \sum_{i=1}^N \Gamma \, S_i \, \rho \, S^\dagger_i + \sum_{j \neq i} \mathrm{Re} \, C_{ij} \, S_i \, \rho \, S^\dagger_i \notag \\
&& \hspace*{-1cm} + \sum_{j \neq i}\sum_{\alpha=x,y,z} \mathrm{Re} \, D_{ij}^\alpha \, \big[ \, S_i \, \rho \, S^\dagger_j \big( B_{j \alpha} + B_{j \alpha}^\dagger \big) \notag \\
&& \hspace*{-1cm} - \big( B_{i \alpha} + B_{i \alpha}^\dagger \big) S_i \, \rho \, S^\dagger_j \, \big] \notag \\
&& \hspace*{-1.2cm} + \sum_{i,j=1}^N\sum_{\alpha,\beta=x,y,z}\hspace*{-0.3cm} \mathrm{Re} \, E_{ij}^{\alpha \beta} \, \big[ \, \big( B_{i \alpha} + B_{i \alpha}^\dagger \big)\big( B_{i \beta} + B_{i \beta}^\dagger \big) S_i \, \rho \, S^\dagger_j  \notag \\
&& \hspace*{-1cm} -2 \big( B_{i \alpha} + B_{i \alpha}^\dagger \big) S_i \, \rho \, S^\dagger_j \big( B_{j \beta} + B_{j \beta}^\dagger \big) \notag \\
&& \hspace*{-1cm} + S_i \, \rho \, S^\dagger_j  \big( B_{j \alpha} + B_{j \alpha}^\dagger \big)\big( B_{j \beta} + B_{j \beta}^\dagger \big) \, \big] \, . ~~~~~
\end{eqnarray}
As in the case of fixed positions, the spontaneous emission rate can be higher (superradiance) or lower (subradiance) than its non-interacting value of $\Gamma$ per excited atom, depending on the distance between the atoms.  With motion included, there is also the possibility that a spontaneous emission creates or destroys phonons, at rates $ \mathrm{Re} \, D_{ij}^\alpha,\, \mathrm{Re} \, E_{ij}^{\alpha\beta}$.  The single particle recoil heating is included as the $i=j$ term $\mathrm{Re} \, E_{ii}^{\alpha \beta}=\frac{1}{10}\eta_\alpha \eta_\beta \Gamma  \, \left(\mathbf{\hat D}_{01 \alpha} \mathbf{\hat D}_{01 \beta}-2\delta_{\alpha\beta} \right)$, where $\delta_{\alpha\beta}$ is the Kronecker delta.

In the limit of infinitely strong confinement ($\eta_\alpha\rightarrow 0$, and hence $D_{ij}^\alpha, \, E_{ij}^{\alpha\beta}\rightarrow 0$), Eqs.~(\ref{Hcond}),~(\ref{reset}) reduce to the known master equation for fixed dipole interacting atoms \cite{two-dipole1,two-dipole2,Agarwal}.

\section{Fourier mode representation} \label{fourier}

The lattice has discrete translational symmetry, and we can hence simplify the above Hamiltonian (\ref{Hcond}) and reset operator (\ref{reset}) by transforming to Fourier space, which we do in this section.

\subsection{Mode operators}

The Fourier mode operators $s_\mathbf{p}$, $b_{\mathbf{p} \alpha}$ are labelled by a wavevector $\mathbf{p}$, which runs over the first Brillouin zone $-\pi/a<p_\alpha\leq\pi/a$, where $a$ is the lattice spacing.  They are defined in terms of the single site electronic excitation and phonon operators $S_i$, $B_{i\alpha}$ by
\begin{eqnarray} \label{sb's}
s_\mathbf{p} &=&\frac{1}{\sqrt{N}}\sum\limits_{i=1}^N\mathrm{e}^{-\mathrm{i}\mathbf{p}\cdot\mathbf{X}_i}S_i \, , \notag \\
b_{\mathbf{p}\alpha} &=&\frac{1}{\sqrt{N}}\sum\limits_{i=1}^N\mathrm{e}^{-\mathrm{i}\mathbf{p}\cdot\mathbf{X}_i}B_{i\alpha}\, ,
\end{eqnarray}
where $N$ is the total number of atoms and we assume periodic boundary conditions. The creation operators $s_\mathbf{p}^\dagger$, $b_{\mathbf{p} \alpha}^\dagger$ are given by the Hermitian conjugates of Eq.~(\ref{sb's}).  As before, $\alpha$ denotes the direction of vibration of the atoms and can assume the values $x$, $y$, and $z$.

These operators create or destroy collective excitations spread over all the atoms.  The internal atomic excitations are hardcore bosons (no more than one can exist on any one atom), and hence $s_\mathbf{p}$, $s_\mathbf{p}^\dagger$ do not exactly obey normal bosonic commutation relations, but in the limit of small excitation density the difference can be neglected \cite{Holstein,Shah}.  (We cannot use the Jordan-Wigner transformation to turn them into fermions, as their number is not conserved.)  The collective phonons described by $b_{\mathbf{p}\alpha}$ are ordinary bosons, as in the deep lattice limit one atom can have any number of phonons. As we shall see below, the electronic excitations propagate by photon exchange, while the phonons do not propagate to zeroth order in $\eta_\alpha$.

\subsection{The conditional Hamiltonian}

The definitions (\ref{sb's}) can be inverted to give the single site operators $S_i$, $B_{i\alpha}$ in terms of the Fourier mode operators $s_\mathbf{p}$, $b_{\mathbf{p} \alpha}$.  We now substitute this into the conditional Hamiltonian Eq.~(\ref{Hcond}), and make a Lamb-Dicke expansion up to second order in $\eta_\alpha$, using Eqs.~(\ref{expansion})-(\ref{Dij}).  Introducing the notation
\begin{eqnarray}
H_\mathrm{cond}&=&H_\mathrm{cond}^{(0)} + H_\mathrm{cond}^{(1)} + H_\mathrm{cond}^{(2)}  \, ,
\end{eqnarray}
where the superscripts indicate the order of each term with respect to $\eta_\alpha$, we obtain
\begin{widetext}
\begin{eqnarray} \label{Hconds}
H_\mathrm{cond}^{(0)} &=&\frac{1}{2} \hbar \sqrt{N} \Omega(s_{\mathbf{k}_\mathrm{L}}+s^\dagger_{\mathbf{k}_\mathrm{L}}) - \sum\limits_\mathbf{p}\frac{\mathrm{i}}{2 } \hbar c(\mathbf{p}) \, s^\dagger_\mathbf{p}s_\mathbf{p} + \sum_{\mathbf{p},\alpha} \hbar \nu_\alpha b^\dagger_{\mathbf{p}\alpha}b_{\mathbf{p}\alpha} \, , \nonumber \\
H_\mathrm{cond}^{(1)} &=& \sum\limits_{\mathbf{p},\alpha} \frac{\mathrm{i} }{2} \, \hbar \eta_\alpha \Omega \, \hat k_{\mathrm{L} \alpha} (b_{\mathbf{p} \alpha}+b^\dagger_{{\mathbf{-p}}\alpha})(s^\dagger_{\mathbf{k}_\mathrm{L}+\mathbf{p}}-s_{\mathbf{k}_\mathrm{L}-\mathbf{p}}) - \sum\limits_{\mathbf{p},\mathbf{q},\alpha} \frac{\mathrm{i}\hbar}{2\sqrt{N}} [d_\alpha(\mathbf{q})-d_\alpha(\mathbf{p}+\mathbf{q})](b_{\mathbf{p}\alpha}+b^\dagger_{-\mathbf{p} \alpha}) s^\dagger_{\mathbf{p}+\mathbf{q}}s_\mathbf{q} \, , \nonumber \\
H_\mathrm{cond}^{(2)} &=& -\sum\limits_{\mathbf{p},\mathbf{q},\alpha,\beta} \frac{\hbar}{4\sqrt{N}} \,  \eta_\alpha\eta_\beta \Omega \, \hat k_{\mathrm{L} \alpha}\hat k_{\mathrm{L} \beta} (b_{\mathbf{p} \alpha}+b^\dagger_{{\mathbf{-p}}\alpha})(b_{\mathbf{q} \beta}+b^\dagger_{{\mathbf{-q}}\beta})(s^\dagger_{\mathbf{k}_\mathrm{L}+\mathbf{p}+\mathbf{q}}+s_{\mathbf{k}_\mathrm{L}-\mathbf{p}-\mathbf{q}})\notag\\
&&+\sum\limits_{\mathbf{p},\mathbf{q},\mathbf{r},\alpha,\beta} -\frac{\mathrm{i}\hbar }{2N} [e_{\alpha\beta}(\mathbf{r}+\mathbf{q})-2e_{\alpha\beta}(\mathbf{r})+e_{\alpha\beta}(\mathbf{r}-\mathbf{p})](b_{\mathbf{p}\alpha}+b^\dagger_{-\mathbf{p} \alpha}) (b_{\mathbf{q}\beta}+b^\dagger_{-\mathbf{q} \beta})s^\dagger_{\mathbf{r}+\mathbf{q}}s_{\mathbf{r}-\mathbf{p}} \, ,
\end{eqnarray}
\end{widetext}
where addition of wavevectors is defined mod the Brillouin zone.  Here the coefficients $c(\mathbf{p})$, $d_\alpha(\mathbf{p})$, $e_{\alpha\beta}(\mathbf{p})$ are given by
\begin{eqnarray} \label{26}
c(\mathbf{p}) &=& \sum\limits_i \mathrm{e}^{- \mathrm{i} \mathbf{p}\cdot\mathbf{X}_i}C(\mathbf{X}_i) \, , \notag \\
d_\alpha(\mathbf{p}) &=& \sum\limits_i \mathrm{e}^{- \mathrm{i} \mathbf{p}\cdot\mathbf{X}_i}D^\alpha(\mathbf{X}_i) \, , \notag \\
e_{\alpha \beta} (\mathbf{p}) &=& \sum\limits_i \mathrm{e}^{- \mathrm{i} \mathbf{p}\cdot\mathbf{X}_i} E^{\alpha \beta} (\mathbf{X}_i) \, .
\end{eqnarray}
These are the discrete Fourier transforms of the constants $C_{ij}$, $D^\alpha_{ij}$ and $E^{\alpha \beta}_{ij}$ in Eqs.~(\ref{Cij2}) and (\ref{Dij}), with the single atom terms $\hbar (\Delta - {\mathrm{i} \over 2} \Gamma)$ absorbed into them by allowing $i=j$ and defining
\begin{eqnarray} \label{C(0)}
C(\mathbf{0}) \equiv \Gamma +2\mathrm{i} \Delta ~~ \mathrm{and}  ~~ D^\alpha(\mathbf{0})\equiv 0 \, .
\end{eqnarray}
They are functions of wavevector $\mathbf{p}$, and also depend on the lattice spacing and orientation; they are plotted against $\mathbf{p}$ in Fig.~\ref{dipolecoolrates4}(d) for a few example settings.

The main advantage of changing into the Fourier space is that the zeroth order term (in $\eta_\alpha$) $C(\mathbf{X})$ in the dipole-dipole interaction becomes the dispersion $c(\mathbf{p})$ of collective electronic excitations, which allows this term to be treated non-perturbatively.  The first and second order terms $d_\alpha(\mathbf{p})$, $e_{\alpha\beta}(\mathbf{p})$, which we do still have to treat perturbatively, describe interaction processes where an electronic excitation emits or absorbs one or two phonons.  Because of the discrete translational symmetry, wavevector (quasimomentum) is conserved mod the Brillouin zone, with the laser contributing its wavevector $\mathbf{k}_\mathrm{L}$ to each excitation it creates.

In the large system limit $N\rightarrow\infty$, these Fourier sums converge in one dimension (except for a logarithmic divergence at the points $p=\pm k_0$) but diverge in two or more dimensions.  While they can still be calculated for finite sizes, in the presence of such strong finite size effects we cannot neglect the difference between periodic and open boundary conditions.  Physically, this is because in 2D and 3D photons have a significant probability of being reabsorbed after travelling a large distance; in an infinite 3D system there would be no outside for photons to escape to, and they would instead mix with the atomic excitations to form two polariton modes \cite{polariton-3dlattice}.  (In the infinite system limit there are also retardation effects neglected by our master equation, but for practical experimental sizes the boundary is reached before these become significant.)  In the following, we therefore restrict ourselves to one-dimensional lattices.

\subsection{The reset operator}

Similarly, the reset operator $\mathcal{R}(\rho)$ in Eq.~(\ref{reset}) can now be expanded as
\begin{eqnarray}
\mathcal{R}(\rho) &=& \mathcal{R}^{(0)}(\rho) + \mathcal{R}^{(1)}(\rho) + \mathcal{R}^{(2)}(\rho),
\end{eqnarray}
with
\begin{widetext}
\begin{eqnarray} \label{Rs}
\mathcal{R}^{(0)}(\rho) &=&\sum\limits_\mathbf{p}\mathrm{Re} \, c(\mathbf{p}) \, s_\mathbf{p}\rho s^\dagger_\mathbf{p} \, , \nonumber \\
\mathcal{R}^{(1)}(\rho) &=& \sum\limits_{\mathbf{p},\mathbf{q},\alpha}\frac{\mathrm{i} }{\sqrt{N}} \left[ \mathrm{Im} \, d_\alpha(\mathbf{q}) \, s_\mathbf{q}\rho s^\dagger_{\mathbf{p}+\mathbf{q}} (b_{\mathbf{p}\alpha}+b^\dagger_{-\mathbf{p}\alpha})  - \mathrm{Im} \, d_\alpha(\mathbf{p}+\mathbf{q}) \, (b_{\mathbf{p}\alpha}+b^\dagger_{-\mathbf{p}\alpha})s_\mathbf{q}\rho s^\dagger_{\mathbf{p}+\mathbf{q}} \right] \, , \notag \\
\mathcal{R}^{(2)}(\rho) &=& \sum\limits_{\mathbf{p},\mathbf{q},\mathbf{r},\alpha,\beta} \frac{ 1}{N} \left[\mathrm{Re} \, e_{\alpha\beta}(\mathbf{r}+\mathbf{q})(b_{\mathbf{p}\alpha}+b^\dagger_{-\mathbf{p} \alpha}) (b_{\mathbf{q}\beta}+b^\dagger_{-\mathbf{q} \beta})s_{\mathbf{r}-\mathbf{p}}\rho s^\dagger_{\mathbf{r}+\mathbf{q}}\nonumber\right.\\&&\hspace*{-1.5cm}\left. -2\mathrm{Re} \, e_{\alpha\beta}(\mathbf{r})(b_{\mathbf{p}\alpha}+b^\dagger_{-\mathbf{p} \alpha})s_{\mathbf{r}-\mathbf{p}}\rho s^\dagger_{\mathbf{r}+\mathbf{q}} (b_{\mathbf{q}\beta}+b^\dagger_{-\mathbf{q} \beta})+\mathrm{Re} \, e_{\alpha\beta}(\mathbf{r}-\mathbf{p})s_{\mathbf{r}-\mathbf{p}}\rho s^\dagger_{\mathbf{r}+\mathbf{q}}(b_{\mathbf{p}\alpha}+b^\dagger_{-\mathbf{p} \alpha}) (b_{\mathbf{q}\beta}+b^\dagger_{-\mathbf{q} \beta})\right] \, .
\end{eqnarray}
\end{widetext}
Because $\mathrm{e}^{ipx}+\mathrm{e}^{-ipx}=2\cos px$ is real and $\mathrm{e}^{ipx}-\mathrm{e}^{-ipx}=2\mathrm{i}\sin px$ is pure imaginary, even functions such as $C_{ij},E_{ij}^{\alpha\beta}$ transform real part to real part ($\mathrm{Re} \, C_{ij}\rightarrow\mathrm{Re} \, c(\mathbf{p})$) while odd functions such as $\mathrm{Re}D^\alpha_{ij}$ transform real part to imaginary part ($\mathrm{Re}D^\alpha_{ij}\rightarrow\mathrm{i} \, \mathrm{Im} \, d_\alpha(\mathbf{p})$). The zeroth order term describes the spontaneous decay of the electronic excitations, with a wavevector-dependent rate because of dipole-dipole interaction.  The first and second order terms are decays that also create or destroy phonons.
\section{Steady state density matrix} \label{stat}
Under the master equation (\ref{master}), the density matrix converges to a steady state, and for cooling we would like it to converge within a reasonable time to a state with the lowest possible temperature, or equivalently phonon number.  In this section, we calculate this steady state using an open system version of Feynman diagrams.

\subsection{Open system Feynman diagrams}

In this subsection, we describe a modified form of Feynman diagrams that can be used on open systems described by a time independent Markovian master equation.  Our diagrams are similar to those of Keldysh \cite{keldysh-orig,keldysh-review}, but our starting point is a master equation rather than a system plus bath Hamiltonian.

\begin{figure}[t]
\includegraphics{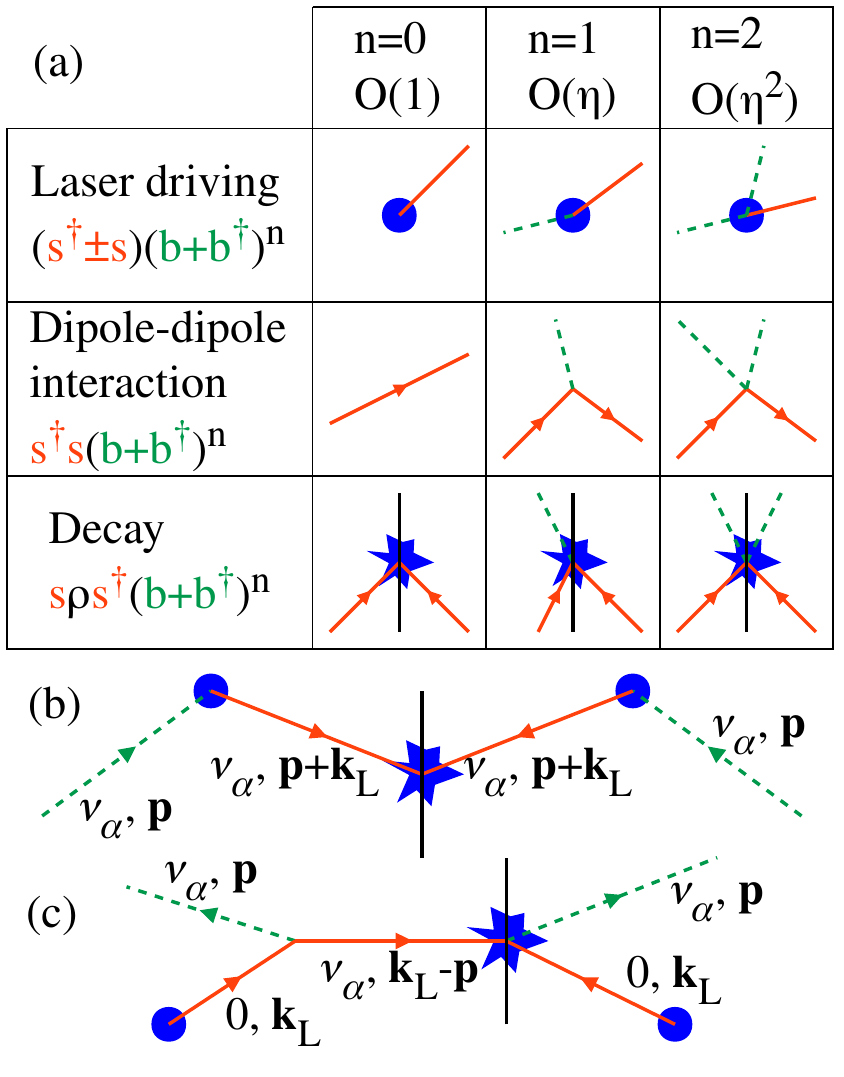}
\vspace*{-0.5cm} \caption{(Color online) Feynman diagrams.  Solid orange line: electronic excitation, dotted green line: phonon, blue circle: laser (de)excitation, blue star: spontaneous decay.  The black vertical line is used to keep track of which operators act from which side of $\rho$.  (a) Table of allowed vertices, by number of phonon lines (order in $\eta_\alpha$) $n$ and type of interaction.  Vertices exist with arbitrarily high $n$, but for our purposes we need only consider up to 2nd order, i.e. $n\leq 2$.  In the case of decay, the operator ordering shown is not the only one allowed: there is always one atomic excitation on each side of $\rho$, but the phonon(s) can be on either or both sides.  Where no arrowhead is shown either propagation direction is allowed.  (b) Example of a rate term (diagram symmetrical about the $\rho$ line), the ordinary sideband cooling process.  (c) Example of an interference term (diagram asymmetrical).  Lines labelled with frequency $E/\hbar$ and wavevector $\mathbf{p}$.} \label{feynman}
\end{figure}

Our diagrams divide the Liouvillean superoperator $\mathcal{L}(\rho)=\dot\rho$ into a propagation part and a vertex part, and treat the vertex part as a perturbation, in the same way as conventional Feynman diagrams divide the Hamiltonian operator \cite{dyson1}.  Propagation terms are those in $H_\mathrm{cond}$ with a matching creation and annihilation operator, i.e. $- \frac{\mathrm{i}}{2} \hbar c(\mathbf{p}) \, s^\dagger_\mathbf{p}s_\mathbf{p}$ for an electronic excitation and $\hbar \nu_\alpha b^\dagger_{\mathbf{p}\alpha}b_{\mathbf{p}\alpha}$ for a phonon.  (These lines represent propagation through time as well as space, so it makes sense for a non-propagating particle such as our phonon to have a propagation line.)  The remaining terms in $H_\mathrm{cond}$ and all terms in $\mathcal{R}(\rho)$ are vertices.  Fig.~\ref{feynman}(a) shows the vertices appearing in our master equation.

Propagation lines are labelled with whatever indices are necessary to fully identify an operator, here $\mathbf{p}$ for an electronic excitation and $\mathbf{p}\alpha$ for a phonon, and also with an energy $E$, which is generally \emph{not} on that particle's dispersion.  Each operator in a vertex must be connected to a matching propagation line, which may go to another vertex or leave the diagram.  The energy $E$ is conserved at vertices (total in=total out), and a particular master equation may have additional conserved quantities, e.g. wavevector if it is translationally invariant (such as ours).  Only particles appearing as operators in the master equation appear as lines in the diagram; particles eliminated during the master equation derivation (here, decay photons) do not.  Classical entities such as the laser field also do not appear as lines, but may contribute momentum, here $\mathbf{k}_\mathrm{L}$ at each laser vertex.

A master equation contains operators acting on both sides of the density matrix $\rho$, and we hence need to keep track of which operators act on the left of $\rho$ and which on the right, which we do with the black vertical line in the diagrams.  Propagation lines and vertices from $H_\mathrm{cond}\rho$ are on the left of this $\rho$ line, while those from $\rho H_\mathrm{cond}^\dagger$ are on the right; a line or vertex on the right hence has the complex conjugate coefficient of an identical one on the left.  Reset operator vertices contain operators acting on both sides of $\rho$, and hence appear only on the $\rho$ line, with propagation lines attached to them from both sides. Propagation lines cannot cross the $\rho$ line without going through a reset vertex.

We choose to make the arrow on a propagation line point forward in time, i.e. from the vertex where that excitation is created to that where it is destroyed.  Since $s^\dagger_\mathbf{p}\ket{0}=\ket{1_\mathbf{p}}$ but $\bra{1_\mathbf{p}}s^\dagger_\mathbf{p}=\bra{0}$, this is the usual creation operator (of the vertex) to annihilation operator on the left side of the diagram, but annihilation operator to creation operator on the right side.  With this convention, energy/momentum conservation at a reset vertex means equal energy/momentum on each side of $\rho$.  (The alternative of creation operator to annihilation operator on both sides would also be self-consistent, and would make the conservation laws in=out everywhere, but would make the arrows point backwards in time on the right.)

To evaluate such a diagram, we replace each vertex by its coefficient (in the master equation, i.e. including the $\pm\mathrm{i}/\hbar$ where applicable) and each internal line by its propagator $1/(-\text{coefficient}\mp\frac{\mathrm{i}}{\hbar}E)$, where the $-$ applies to the left side of $\rho$ and the $+$ to the right.  Lines entering or leaving the diagram have on-resonance energy.  As in conventional Feynman diagrams, for tree diagrams the external lines' energies (and momenta, if conserved) determine all the internal ones by conservation, while for diagrams containing loops it is necessary to integrate over the undetermined energies and momenta.  For example, ordinary sideband cooling is represented by the diagram Fig.~\ref{feynman}(b), which has the value

\begin{eqnarray}
&&\left(\frac{\eta_\alpha\Omega\hat k_{\mathrm{L}\alpha}}{2}\right)\left(\frac{1}{\frac{1}{2}c(\mathbf{p}+\mathbf{k}_\mathrm{L})-\mathrm{i} \nu_\alpha}\right)\mathrm{Re}\,c(\mathbf{p}+\mathbf{k}_\mathrm{L})\nonumber\\
&&\times\left(\frac{1}{\frac{1}{2}c^*(\mathbf{p}+\mathbf{k}_\mathrm{L})+\mathrm{i} \nu_\alpha}\right)\left(\frac{\eta_\alpha\Omega\hat k_{\mathrm{L}\alpha}}{2}\right)\nonumber\\
&&=\frac{\left(\eta_\alpha\Omega\hat k_{\mathrm{L}\alpha}\right)^2\mathrm{Re}\,c(\mathbf{p}+\mathbf{k}_\mathrm{L})}{|c(\mathbf{p}+\mathbf{k}_\mathrm{L})-2\mathrm{i} \nu_\alpha|^2}\, ,
\end{eqnarray}
where $\mathbf{p}$ is the phonon momentum (which is the same on both sides by conservation).

The resulting value is that diagram's contribution to the transition rate (not amplitude) between its start and end states, which are given by the particles entering and leaving it.  Symmetrical diagrams such as Fig.~\ref{feynman}(b) give the rate of the process appearing on both sides, while asymmetrical diagrams such as Fig.~\ref{feynman}(c) give the interference term between their left and right processes.  In the case of coherent processes (i.e. not involving any reset terms), the left and right sides are disconnected, and the transition rate is simply the product of the left and right amplitudes.  As already explained, amplitudes on the right are complex conjugated, so this is consistent with $\text{rate}=|\text{amplitude}|^2$.

\subsection{Heating and cooling rates}

In this section we use these diagrams to calculate the heating and cooling rates.

\begin{figure}[t]
\includegraphics{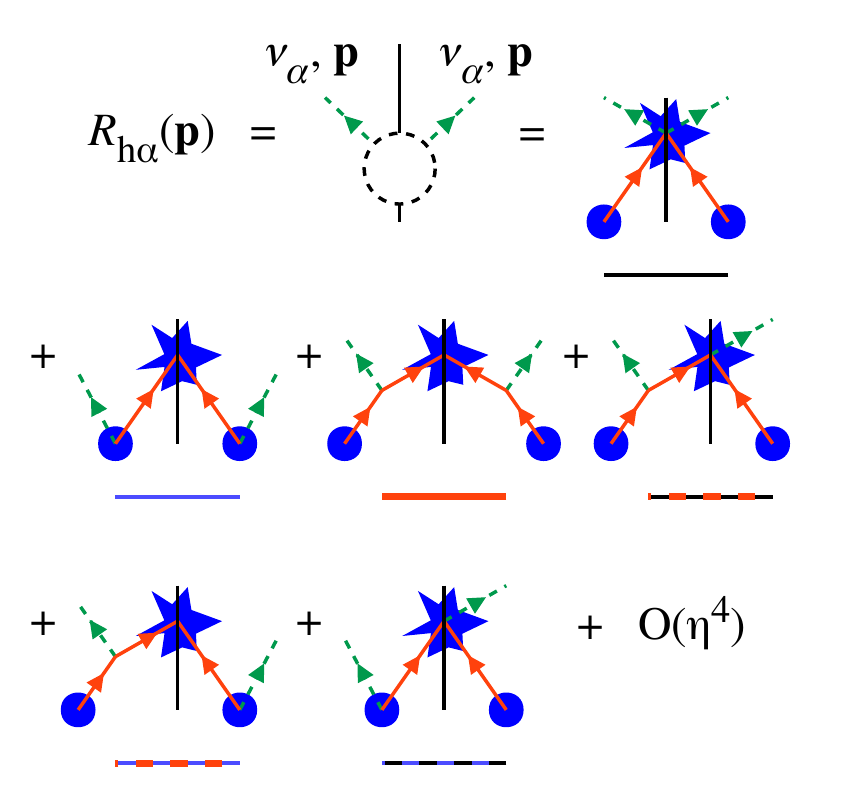}
\vspace*{-0.5cm} \caption{(Color online) Feynman diagrams for the heating rate $R_{\mathrm{h}\alpha}(\mathbf{p})$, as evaluated in Eq.~\ref{feynmansum}.  The cooling rate diagrams are the same except that all phonon arrows are reversed.  As in Fig.~\ref{feynman}, solid orange lines are atomic excitations, dotted green lines are phonons, blue circles are laser driving, blue stars are spontaneous decay.  The black dotted circle denotes any allowed diagram with the given external lines.  The line under each diagram is the color used to plot its value in Fig.~\ref{dipolecoolrates4}(b).  The symmetrical diagrams (first 3) are rate terms; the asymmetrical diagrams (last 3) are interference terms, and are to be read as including both mirror-image versions. \label{feynman3}}
\end{figure}

The lowest order (in the Lamb-Dicke parameters $\eta_\alpha$) processes that change the phonon number increase or decrease it by 1 on both sides of $\rho$, as changing it on only one side is forbidden by energy conservation.  This involves 2 external phonon lines (one on each side) which must both end on vertices, and is hence at least 2nd order in $\eta_\alpha$.  In each case there are 6 such diagrams which must be summed to obtain the rate, shown in Fig.~\ref{feynman3} for creating a phonon (heating); for destroying a phonon (cooling), they are the same except with all phonon lines reversed.  For simplicity we restrict to the fully anisotropic case (three different phonon frequencies $\nu_x$, $\nu_y$, $\nu_z$), where energy conservation requires that $\alpha$ be the same on both sides.

Evaluating these diagrams gives
\begin{eqnarray}
&&R_{\mathrm{h}\alpha}(\mathbf{p}),\,R_{\mathrm{c}\alpha}(\mathbf{p})=\frac{-2\Omega^2\mathrm{Re}\,e_{\alpha\alpha}(\mathbf{k}_\mathrm{L}\mp\mathbf{p})}{|c(\mathbf{k}_\mathrm{L})|^2}\notag\\
&&+\frac{\left(\eta_\alpha\Omega\hat k_{\mathrm{L}\alpha}\right)^2\mathrm{Re}\,c(\mathbf{k}_\mathrm{L}\mp\mathbf{p})}{|c(\mathbf{k}_\mathrm{L}\mp\mathbf{p})\pm 2\mathrm{i} \nu_\alpha|^2}\notag\\
&&+\frac{\Omega^2|d_\alpha(\mathbf{k}_\mathrm{L})-d_\alpha(\mathbf{k}_\mathrm{L}\mp\mathbf{p})|^2\mathrm{Re}\,c(\mathbf{k}_\mathrm{L}\mp\mathbf{p})}{|c(\mathbf{k}_\mathrm{L})|^2|c(\mathbf{k}_\mathrm{L}\mp\mathbf{p})\pm 2\mathrm{i} \nu_\alpha|^2}\notag\\
&&+2\mathrm{Re}\left\{-\frac{\mathrm{i}\Omega^2[d_\alpha(\mathbf{k}_\mathrm{L})-d_\alpha(\mathbf{k}_\mathrm{L}\mp\mathbf{p})]\mathrm{Im}\,d_\alpha(\mathbf{k}_\mathrm{L}\mp\mathbf{p})}{|c(\mathbf{k}_\mathrm{L})|^2[c(\mathbf{k}_\mathrm{L}\mp\mathbf{p})\pm 2\mathrm{i} \nu_\alpha]}\right.\notag\\
&&+\frac{\mathrm{i}\Omega^2\eta_\alpha\hat k_{\mathrm{L}\alpha}[d_\alpha(\mathbf{k}_\mathrm{L})-d_\alpha(\mathbf{k}_\mathrm{L}\mp\mathbf{p})]\mathrm{Re}\,c(\mathbf{k}_\mathrm{L}\mp\mathbf{p})}{c(\mathbf{k}_\mathrm{L})|c(\mathbf{k}_\mathrm{L}\mp\mathbf{p})\pm 2\mathrm{i} \nu_\alpha|^2}\notag\\
&&-\left.\frac{\Omega^2\eta_\alpha\hat k_{\mathrm{L}\alpha}\mathrm{Im}\,d_\alpha(\mathbf{k}_\mathrm{L}\mp\mathbf{p})}{c^*(\mathbf{k}_\mathrm{L})[c(\mathbf{k}_\mathrm{L}\mp\mathbf{p})\pm 2\mathrm{i} \nu_\alpha]}\right\}+O(\eta^4)\, ,\label{feynmansum}
\end{eqnarray}
where $\mathbf{p}$ is the phonon momentum, which must be the same on both sides of $\rho$ by conservation, and the upper sign applies to phonon creation (i.e. heating) and the lower sign to phonon destruction (cooling).  The $2\mathrm{Re}$ on the 3 asymmetric diagrams is because we also need to include their mirror images, which have the complex conjugate value.

At the same order in the $\eta_\alpha$, there are also a self-energy correction $\Sigma_\alpha(\mathbf{p})$ to the phonon line given by diagrams with one ingoing and one outgoing phonon line on the same side of $\rho$, and a vacuum term $V$ given by diagrams with no external lines.  These are evaluated in the Appendix (Eqs.~(\ref{selfenergy}), (\ref{vacbubble})).  Together, these give an effective master equation for the phonons,
\begin{eqnarray}\label{fmaster}
\dot\rho&=&V\rho+\sum\limits_{\mathbf{p},\alpha}\left[(\Sigma_\alpha(\mathbf{p})-\mathrm{i}\nu_\alpha)b^\dagger_{\mathbf{p}\alpha}b_{\mathbf{p} \alpha}\, \rho  \right.\notag\\
&&+(\Sigma^*_\alpha(\mathbf{p})+\mathrm{i}\nu_\alpha)\rho\, b^\dagger_{\mathbf{p}\alpha}b_{\mathbf{p} \alpha}+R_{\mathrm{h}\alpha}(\mathbf{p})b^\dagger_{\mathbf{p} \alpha}\, \rho \,b_{\mathbf{p} \alpha}\notag \\
&&+\left.R_{\mathrm{c}\alpha}(\mathbf{p})b_{\mathbf{p} \alpha}\, \rho \,b^\dagger_{\mathbf{p} \alpha} \right] \, .
\end{eqnarray}
This conserves probability as $\mathrm{Re}\,\Sigma_\alpha(\mathbf{p})=-\frac{1}{2}(R_{\mathrm{h}\alpha}(\mathbf{p})+R_{\mathrm{c}\alpha}(\mathbf{p})$ and $V=-\sum_{\mathbf{p},\alpha}R_{\mathrm{h}\alpha}$.  It also agrees with the steady state equation (\ref{43}) calculated by direct perturbative expansion in the Appendix.
\subsection{Steady state phonon distribution}

We now use these equations to calculate the steady state of the phonons.  We also introduce the mean phonon number as a measure of heat content, as the steady state does not have a definite temperature.

Eq.~(\ref{fmaster}) does not couple different phonon modes $\mathbf{p} \alpha$, so the phonon steady state $\rho^{(0)}_\mathrm{phn}$ is a product state over modes.  As $\mathrm{Re}\, \Sigma_\alpha(\mathbf{p})<0$, phonon coherences decay, so the steady state is a mixture (not a superposition) over phonon numbers, 
\begin{eqnarray} \label{42}
\rho^{(0)}_\mathrm{phn} &=& \prod_{\mathbf{p},\alpha} \sum_{m_{\mathbf{p} \alpha}} c_{m_{\mathbf{p} \alpha}} \, |m_{\mathbf{p} \alpha} \rangle \langle m_{\mathbf{p} \alpha} | \, .
\end{eqnarray}
Here the coefficient $c_{m_{\mathbf{p} \alpha}} $ is the probability of having $m_{\mathbf{p} \alpha}$ phonons in the mode $\mathbf{p} \alpha$.  As the state is a product state, the phonon distributions of different modes are independent.

Substituting (\ref{42}) into (\ref{fmaster}) and setting $\dot\rho=0$, we obtain
\begin{eqnarray} 
0 &=& m_{\mathbf{p} \alpha} \, R_{\mathrm{h}\alpha} (\mathbf{p}) \, c_{m_{\mathbf{p} \alpha} - 1}  \notag \\
&& \hspace*{-0.3cm} - \left[ (m_{\mathbf{p} \alpha} +1) R_{\mathrm{h}\alpha} (\mathbf{p}) 
+ m_{\mathbf{p} \alpha}R_{\mathrm{c} \alpha} (\mathbf{p}) \right] \, c_{m_{\mathbf{p} \alpha}} \notag \\
&& \hspace*{-0.3cm} + (m_{\mathbf{p} \alpha} + 1) \,  R_{\mathrm{c}\alpha}(\mathbf{p}) \, c_{m_{\mathbf{p} \alpha} + 1} \, .
\end{eqnarray}
One solution of this equation is
\begin{eqnarray} \label{47}
 c_{m_{\mathbf{p} \alpha} } &=& \left[ {R_{\mathrm{h} \alpha} (\mathbf{p}) \over R_{\mathrm{c}\alpha} (\mathbf{p}) }\right]^{m_{\mathbf{p} \alpha} } \, , 
\end{eqnarray}
and it is unique up to normalization because once $c_{0}$ is chosen the others can be determined iteratively. 

As expected for a steady state, heating and cooling are in detailed balance: the transition rate from $m$ to $m+1$ phonons is equal to the transition rate from $m+1$ to $m$. A normalizable steady state only exists for the modes where the heating rate $R_{\mathrm{h} \alpha} (\mathbf{p})$ is smaller than the corresponding cooling rate $R_{\mathrm{c} \alpha} (\mathbf{p})$.  If not, we obtain unbounded heating.

We finally consider how to measure the heat content of this steady state, as in the next section we will want to minimise this.  For each individual phonon mode $\mathbf{p}\alpha$, the state (\ref{47}) is a thermal state with temperature
\begin{eqnarray} 
T_{\mathbf{p} \alpha} &=& \frac{\nu_{\alpha}}{k_\mathrm{B} \, \log \left[R_{\mathrm{c}\alpha}(\mathbf{p})/R_{\mathrm{h} \alpha} (\mathbf{p}) \right] } \, .
\end{eqnarray}
However, the phonon state as a whole is not thermal because different modes $\mathbf{p}\alpha$ have different temperatures.  We therefore instead use the mean phonon number as our measure of heat content, as this is defined for any finite energy state, and when multiplied by $\hbar\nu_\alpha$ gives the mean phonon energy.  For a single mode it is
\begin{eqnarray}
\langle m_{\mathbf{p} \alpha}  \rangle_\mathrm{ss} &=& {\sum\limits_{m_{\mathbf{p} \alpha}}  m_{\mathbf{p} \alpha} \, c_{m_{\mathbf{p} \alpha} }
\over \sum\limits_{m_{\mathbf{p} \alpha}} c_{m_{\mathbf{p} \alpha} }}\nonumber\\
& =& {R_{\mathrm{h}\alpha} (\mathbf{p}) \over R_{\mathrm{c} \alpha} (\mathbf{p}) - R_{\mathrm{h}\alpha} (\mathbf{p})} \, .
\end{eqnarray}
For multiple modes it can then be obtained by simple averaging; in particular, the mean phonon number per atom over all wavevectors is
\begin{eqnarray}
\langle m\rangle_\mathrm{ss}=\frac{1}{N}\sum_\mathbf{p} \langle m_{\mathbf{p} \alpha}  \rangle_\mathrm{ss} \, .
\end{eqnarray}
We do not average over motion directions $\alpha$, as in a single laser setup we expect cooling in one dimension (along the laser) and heating in the other two.

\section{Examples and parameter optimization}\label{examples}

In this section we evaluate the steady state phonon number for various parameter values, and in particular consider what settings minimize it.  We find that the dipole-dipole interaction has a significant effect on the steady state phonon number, at lattice spacing $a=\lambda/2$ typically $\sim 100\%$ at the most strongly affected phonon wavevectors and $\sim 20\%$ averaged over all wavevectors.  This effect can have either sign depending on the parameters; in particular, there are settings that give a lower phonon number $\langle m\rangle_\mathrm{ss}$ than \emph{any} non-interacting settings with the same lattice strength $\nu$.

\subsection{Ordinary (non-interacting) sideband cooling}

As a check on our calculation we first consider the case of atoms far enough apart to neglect their dipole-dipole interaction. In this case, we should obtain ordinary sideband cooling of each atom independently. No interaction means 
\begin{eqnarray}
&& C_{ij} = 0, ~D_{ij}^\alpha = 0,~ E_{ij}^{\alpha \beta} = 0 ~~ \forall ~ i\neq j \, .
\end{eqnarray}
This leaves only the single atom terms $C_{ii}$ (Eq.~ (\ref{C(0)})) and $E_{ii}^{\alpha\beta}$, as $D_{ii}^\alpha=0$ by definition (cf.~Eq.~ (\ref{C(0)})).  The Fourier transforms are hence
\begin{eqnarray}
c(\mathbf{p}) =& C_{ii}=&\Gamma +2\mathrm{i} \, \Delta \, , ~~ d_\alpha(\mathbf{p}) = 0 \, , \notag \\
e_{\alpha \beta} (\mathbf{p}) =&E_{ii}^{\alpha\beta}=& {\eta_\alpha \eta_\beta \Gamma \over 10} \, \left(\mathbf{\hat D}_{01 \alpha}\mathbf{\hat D}_{01 \beta}-2\delta_{\alpha\beta} \right) \, ,
\end{eqnarray}
independent of $\mathbf{p}$, where $\mathbf{\hat D}_{01 \alpha}$ is the component of the dipole unit vector $\mathbf{\hat D}_{01}$ along $\alpha$.

In ordinary sideband cooling, the cooled motion $\alpha$ is parallel to the laser wave vector ($\vartheta=0$) and hence $\hat k_{\mathrm{L} \alpha} =1$ and $\mathbf{\hat D}_{01 \alpha} = 0$ for this motion.  Substituting into Eq.~(\ref{feynmansum}), we obtain the heating and cooling rates
\begin{eqnarray} \label{M3}
R_{\mathrm{h} \alpha} &=& \eta_\alpha^2 \Omega^2 \Gamma \, \left[ \frac{1}{\Gamma^2 +4 ( \Delta + \nu_\alpha)^2} + \frac{2}{5(\Gamma^2 + 4 \Delta^2)} \right] \, , \notag \\
R_{\mathrm{c} \alpha} &=& \eta_\alpha^2 \Omega^2 \Gamma \, \left[ \frac{1}{\Gamma^2 +4 ( \Delta - \nu_\alpha)^2} + \frac{2}{5(\Gamma^2 + 4 \Delta^2)} \right] \, , \notag \\
\end{eqnarray}
independent of $\mathbf{p}$.  These are the standard results for single particle sideband cooling \cite{laser-cooling-theory,ion-cooling-review}.

In this case, only the first 2 diagrams in Fig.~\ref{feynman3} are nonzero.  The first term in each of these equations represents a phonon being created (heating) or destroyed (cooling) when the laser excites an atom, followed by spontaneous decay with no change in phonon state.  Cooling is resonantly enhanced when $\Delta\approx\nu_\alpha$, as expected from Fig.~\ref{dipolecoollevels}.  The second term represents laser excitation with no change in phonon state, followed by spontaneous decay with phonon creation or destruction.  This is recoil heating; as it has equal heating and cooling rates, it alone would have an infinite equilibrium temperature, so it has a heating effect on any finite temperature state.

\begin{figure}[t]
\includegraphics{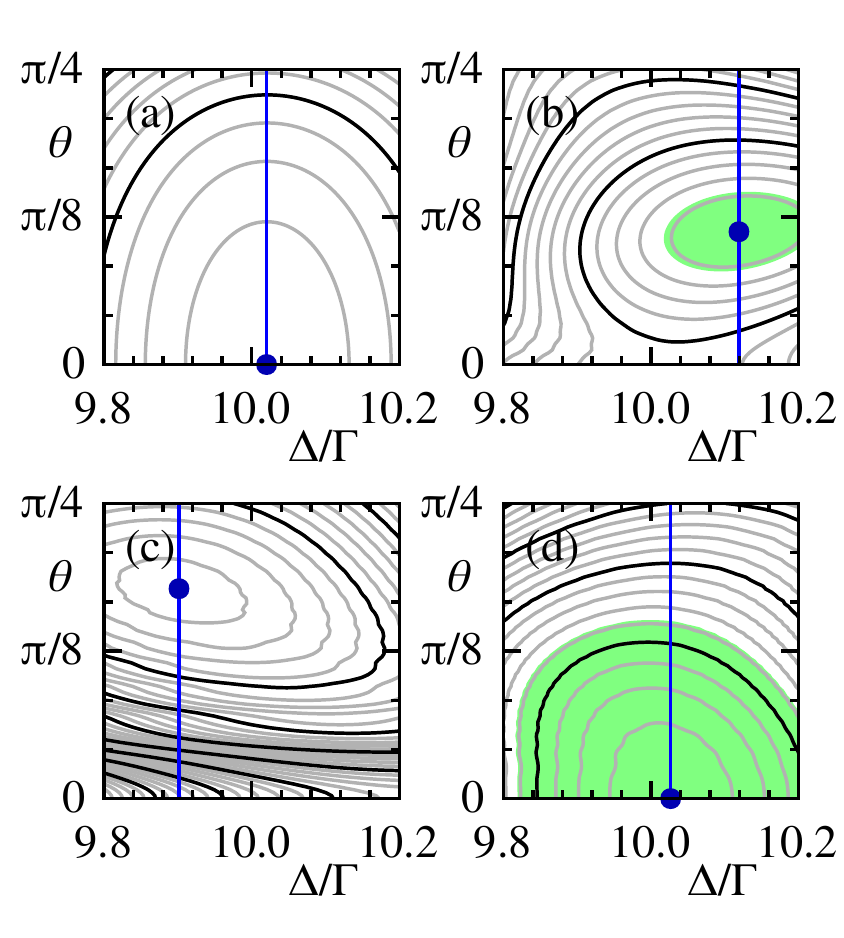}
\vspace*{-0.5cm} \caption{(Color online) Contour plots of the steady state phonon number $\langle m\rangle_\mathrm{ss}$, averaged over wavevectors, as a function of the laser detuning $\Delta$ and the angle $\vartheta$ between the atomic motion and the laser direction. (a) Non-interacting case.  (b) Across-lattice motion, $a=0.5\lambda$.  (c) Along-lattice motion, $a=0.5\lambda$.  (d) Along-lattice motion, $a=0.744\lambda$.  In all cases $\nu = 10\Gamma$ and the atomic dipole moment $\mathbf{D}_{01}$ is perpendicular to the wavevector $\mathbf{k}_\mathrm{L}$ of the incoming laser.  All cases are symmetric about $\vartheta=0$; to save space, only positive $\vartheta$ are shown here.  The green shaded areas are where the phonon number is lower than its minimum non-interacting value.  The blue dots are the minima, and the blue lines the sections plotted in Fig.~\ref{thetaplot}.  The contour interval is $10^{-4}$ phonons per atom.} \label{dipolecoolcontour}
\end{figure}

Fig.~\ref{dipolecoolcontour}(a) is a contour plot of the steady state phonon number $\langle m\rangle_\mathrm{ss}$ as a function of the laser detuning $\Delta$ and the angle $\vartheta$ between the cooling laser and the motion (cf.~Fig.~\ref{dipolecoollayout}).  As expected, the lowest steady state phonon number is obtained for $\vartheta=0$ and $\Delta$ close (but not exactly equal) to the phonon frequency $\nu_\alpha$.

Since we consider a one laser setup, the atomic motion $\beta$ orthogonal to the laser wave vector is not cooled, but is still subject to recoil heating. For this motion we have $\hat k_{\mathrm{L} \beta} =0$, and hence 
\begin{eqnarray} 
R_{\mathrm{h} \beta} = R_{\mathrm{c} \beta} = \frac{\eta_\beta^2 \Gamma \Omega^2}{5(\Gamma^2 + 4 \Delta^2)} \, \left(2-\mathbf{\hat D}_{01 \beta}^2 \right) \, .
\end{eqnarray}
Both the cooling and the heating rate are the same for this motion. This means the mean phonon number $\langle m\rangle_\mathrm{ss}$ for this motion diverges, i.e. unbounded heating, as expected. A diagonally oriented laser could cool all three dimensions, but inefficiently as it can only be in resonance with one of them.  Alternatively, three lasers could be used, one along each dimension.

\subsection{Interacting case: results}

\begin{figure}[t]
\includegraphics{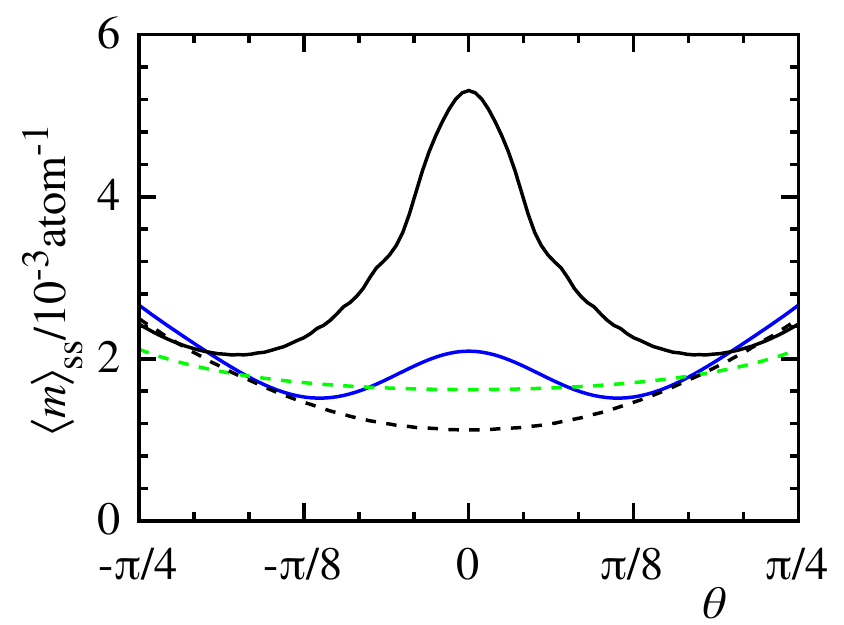}
\vspace*{-0.5cm} \caption{(Color online) Steady state phonon number $\langle m\rangle_\mathrm{ss}$, averaged over wavevectors, against $\vartheta$.  The detuning $\Delta$ is in each case chosen so the plot passes through the minimum of phonon number (forming the sections of Fig.~\ref{dipolecoolcontour} along the blue lines).  Dotted green: no interaction ($a=\infty$).  Solid blue: across-lattice motion, $a=0.5\lambda$.  Solid black: along-lattice motion, $a=0.5\lambda$.  Dotted black: along-lattice motion, $a=0.744\lambda$.} \label{thetaplot}
\end{figure}

\begin{figure}[t]
\includegraphics{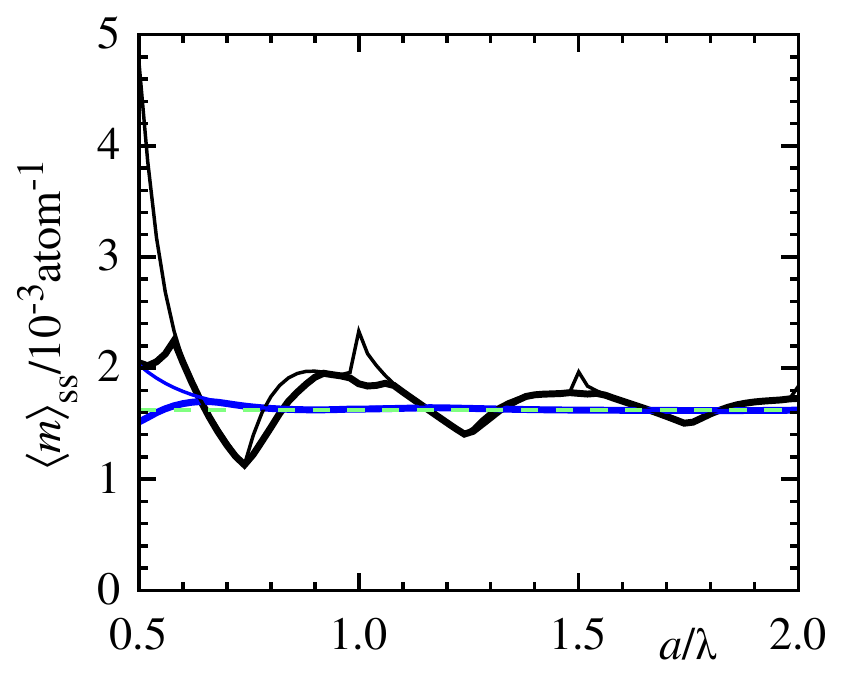}
\vspace*{-0.5cm} \caption{(Color online) Steady state phonon number $\langle m\rangle_\mathrm{ss}$, averaged over wavevectors, against lattice spacing $a$, for along-lattice motion (black) and across-lattice motion (blue).  Thin lines are standard sideband cooling $\Delta=\nu,\vartheta=0$, thick lines are $\Delta,\vartheta$ chosen to minimize $\langle m\rangle_\mathrm{ss}$.  The dotted horizontal line is the non-interacting minimum.\label{occminsp}}
\end{figure}

We now move to the case with dipole-dipole interaction, and consider varying the lattice spacing $a$ and the laser detuning $\Delta$ and orientation $\vartheta$, for both the along- and across-lattice modes (as defined in Fig.~\ref{dipolecoollayout}).  We restrict to the case where the laser wavevector $\mathbf{k}_\mathrm{L}$ is orthogonal to the atomic dipole moment $\mathbf{D}_{01}$, which requires the least laser power for a given $\Omega$, and both are in the $x$\textendash$y$ plane.

Figs.~\ref{dipolecoolcontour}(b,c,d) are contour plots of the steady state phonon number $\langle m\rangle_\mathrm{ss}$ against the laser detuning $\Delta$ and angle $\vartheta$, for three choices of lattice spacing and orientation.  Fig.~\ref{thetaplot} plots the $\Delta=\mathrm{constant}$ cross-section through the minimum of $\langle m\rangle_\mathrm{ss}$, for the same cases.  Fig.~\ref{occminsp} plots steady state phonon number against lattice spacing.

The numerical results presented here were all obtained at $N=200$ with the dipole interaction cut off at that range (i.e. each pair of particles have one interaction term, not an infinite series of image interaction terms) and $\nu=10\Gamma$.  We have tried other $N,\nu$ and found the same qualitative results for $N\gtrsim 5$ and $\nu\gtrsim 5\Gamma$ (along-lattice motion) or $\gtrsim 2\Gamma$ (across-lattice motion).  Increasing $\nu$ decreases the steady state phonon number, for large $\nu/\Gamma$ as approximately $\langle m\rangle_\mathrm{ss}\sim (\Gamma/\nu)^2$, as in the non-interacting case.

We first consider across-lattice motion with lattice spacing $a=0.5\lambda$ (Fig.~\ref{dipolecoolcontour}(b)).  If we leave the settings at the non-interacting optimum of $\vartheta=0,\Delta=10.02\Gamma$, the interaction increases the steady state phonon number from $1.62\times 10^{-3}$ to $2.03\times 10^{-3}$ per atom.  However, if we re-optimize $\vartheta,\Delta$ for the interacting system, we can reach a minimum of $1.51\times 10^{-3}$ phonons per atom at $\vartheta=\pm 0.35,\Delta=10.12\Gamma$.  This is lower than the non-interacting minimum, i.e. dipole interaction can \emph{improve} cooling effectiveness.  In this case, this requires a non-zero angle $\vartheta$ between the laser and the motion to be cooled.

For along-lattice motion, the effect of interaction is stronger, as the longest range term of $C(\mathbf{x})$ (Eq.~(\ref{Cij}), Fig.~\ref{Cplot}) acts across the dipole, which with our assumptions includes along the laser.  At lattice spacing $a=0.5\lambda$ (Fig.~\ref{dipolecoolcontour}(c)), there is a high peak at $\vartheta=0$ ($4.76\times 10^{-3}$ phonons per atom at $\vartheta=0,\Delta=10.02\Gamma$), which slowly gets higher as $N$ increases.  This can be avoided by using a non-zero $\vartheta$, with a minimum of $2.05\times 10^{-3}$ per atom at $\vartheta=\pm 0.56,\Delta=9.90\Gamma$.  While worse than the non-interacting case, this is much better than not optimizing.

If the lattice spacing is allowed to vary (which can be done by using lattice lasers crossed at an angle instead of head-on), it again becomes possible to reach phonon numbers below the non-interacting minimum.  The minimum of $1.14\times 10^{-3}$ per atom is reached at $a=0.744\lambda,\vartheta=\pm 0.05,\Delta=10.04\Gamma$ (Fig.~\ref{dipolecoolcontour}(d)).  There are also local minima at $a=1.248\lambda$ and $a=1.749\lambda$ (Fig.~\ref{occminsp}), which are less deep but still go below the non-interacting minimum.

In all three interacting cases, the phonon number increases a few times faster than in the non-interacting case as we move away from the optimum $\vartheta$.  However, $\pm 5^\circ$ is still sufficient accuracy for better than non-interacting cooling.

The small waves in the contours are $N$-dependent, decreasing in both amplitude and wavelength as $N$ increases.  We conjecture that they are an artifact of the periodic boundary conditions.

We do not consider lattice spacings less than half a wavelength, as the Brillouin zone then contains wavevectors $|\mathbf{p}|>k_0$, at which $\mathrm{Re}\,c(\mathbf{p})$ is very small (zero in the $N\rightarrow\infty$ limit).  Such slowly decaying excitations are additional $\mathcal{L}_0\approx 0$ states, so our perturbative expansion is not valid in this regime.
\subsection{Interacting case: analysis}

\begin{figure*}[t]
\includegraphics{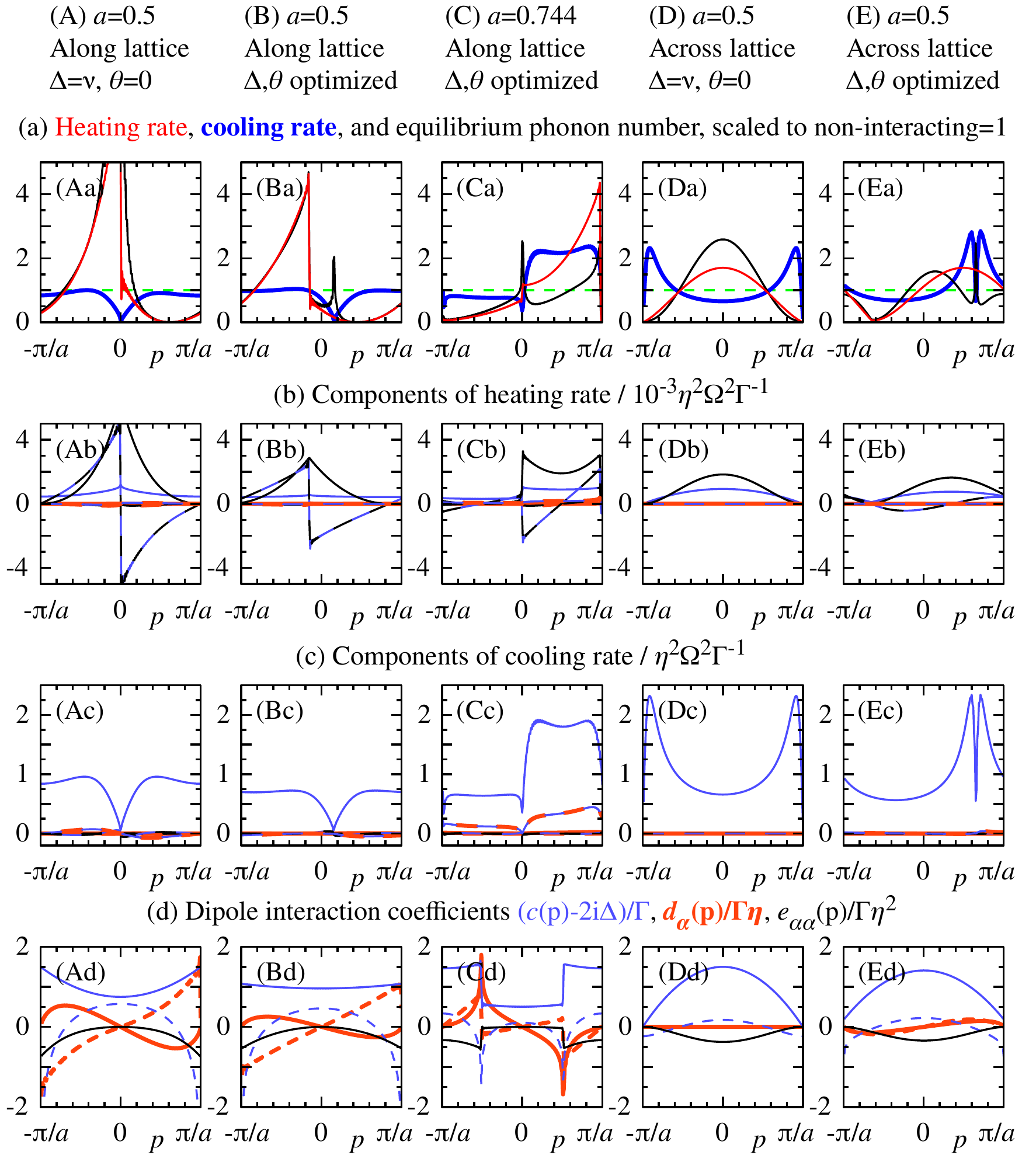}
\caption{(Color online) Various functions of the (one-dimensional) phonon wave vector $p$.  Different rows are different functions, different columns are different settings, given at the top (lattice spacing $a$, cooling of either the along- or across-lattice motion, laser detuning and direction either standard ($\Delta=\nu$, $\vartheta=0$) or optimized (for minimum total phonon number)). (a) Heating rate $R_{\mathrm{h}\alpha}(\mathbf{p})$ (thin red), cooling rate $R_{\mathrm{c}\alpha}(\mathbf{p})$ (thick blue) and steady state phonon number (thin black), each scaled to its own non-interacting value (dotted horizontal line).  (b)/(c) Rates of the individual processes contributing to heating/cooling (cf.~Fig.~\ref{feynman3}): emission/absorption of a phonon either when an excitation is created by the laser (thin blue), when it hops by dipole-dipole interaction (thick orange), or when it decays (thin black), and interferences between these (dotted). (d) Fourier space dipole interaction parameters $c(\mathbf{p})$ (thin blue), $d_\alpha(\mathbf{p})$ (thick orange), $e_{\alpha \alpha} (\mathbf{p})$ (thin black), real parts solid, imaginary parts dotted.  In all cases $\nu = 10 \, \Gamma$ and $\alpha$ is the direction of motion being cooled ($x$ for along-lattice or $y$ for across-lattice).   \label{dipolecoolrates4}}
\end{figure*}

To study how this enhanced cooling happens, we now break down the phonon number into simpler parts, which we plot in Fig.~\ref{dipolecoolrates4}.  The first step in this process is to break the wavevector-averaged $\langle m\rangle_\mathrm{ss}$ into the individual modes $\langle m_{\mathbf{p} \alpha}  \rangle_\mathrm{ss}$, so all these plots are against wavevector.  The rows (a)-(d) are the different parts, while the columns (A)-(E) are the 5 settings that we consider here, given at the top of their column.

Fig.~\ref{dipolecoolrates4}(a) plots the single mode steady state phonon number $\langle m_{\mathbf{p} \alpha}  \rangle_\mathrm{ss}$, heating rate $R_{\mathrm{h}\alpha} (\mathbf{p})$ and cooling rate $R_{\mathrm{c}\alpha} (\mathbf{p})$, each of which is scaled so it would be 1 with no dipole-dipole interaction and the same $\vartheta,\Delta$.  We see that all of these are strongly wavevector dependent, with all 5 settings having both regions that are better than the non-interacting value and regions that are worse.  The unoptimized cases (A),(D) have strong heating coinciding in wavevector space with weak cooling, while the optimized cases (B),(C),(E) mostly pair strong heating with strong cooling, which reduces the total phonon number (if $H>h$ and $C>c$, then $H/c+h/C>H/C+h/c$).

Fig.~\ref{dipolecoolrates4}(b) separates the heating rate into the 6 individual diagrams in Fig.~\ref{feynman3}, while Fig.~\ref{dipolecoolrates4}(c) does the same for the cooling rate.  The dominant cooling process in all 5 cases is phonon absorption on laser excitation (blue line in Fig.~\ref{dipolecoolrates4}(c)), i.e. ordinary sideband cooling, as this is resonantly enhanced.  The strongest heating process is usually phonon emission on decay (recoil heating, black line), but emission on laser excitation (blue) and the interference between them (dashed blue/black) are often also significant.  In all these cases, the heating and cooling rates, and hence the time required to reach the steady state, are of the same order of magnitude as the corresponding non-interacting values.

Finally, Fig.~\ref{dipolecoolrates4}(d) plots the Fourier space dipole interaction coefficients.  One effect of interaction is that the cooling process cannot be exactly on resonance at all wavevectors, as the atomic excitations have a non-trivial dispersion $\mathrm{Im}\,c(\mathbf{p})$ while the phonons have a single frequency $\nu$.  In particular, the dips in cooling rate at $\mathbf{p}+\mathbf{k}_\mathrm{L}=\pm k_0$, in all cases except $\vartheta=0$ across-lattice (D), occur because $\mathrm{Im}\,c(\mathbf{p}+\mathbf{k}_\mathrm{L})$ is large (divergent as $N\rightarrow\infty$) there.  (At $a=0.5\lambda$ (A),(B),(E), these are the same point mod the Brillouin zone and there is hence one dip; at other $a$ (C) there are two.)  The along-lattice locally optimal spacings (Fig.~\ref{occminsp}) $a=0.744\lambda$, $a=1.248\lambda$ and $a=1.749\lambda$ are all close to zeros of $\mathrm{Im}\,C(\mathbf{X})$ (Fig.~\ref{Cplot}(b)), which reduce this effect by removing the nearest neighbor (normally largest) term, reducing the size of these dips.

These dips are accompanied by a sudden change in the decay rate $\mathrm{Re}\,c(\mathbf{p})$.  The optimal spacings are also all near an odd number of quarter wavelengths, where $\pm k_0=\pm \pi/2a$ (not necessarily respectively) mod the Brillouin zone, dividing it into a low $\mathrm{Re}\,c(\mathbf{p})$ half and a high $\mathrm{Re}\,c(\mathbf{p})$ half.  This gives them a particularly good matching of strong cooling to strong heating ($a=0.744\lambda$ shown in Fig.~\ref{dipolecoolrates4}(C), the other two are similar), which further reduces their average phonon number.

\section{Conclusions} \label{conc}

To summarize, we have considered resolved sideband cooling in a deep 1D optical lattice, and in particular the effect of the dipole-dipole interaction that exists between any dipole radiating particles.

At the typical optical lattice spacing of half a wavelength, this interaction is not a small perturbation, but is of the same order of magnitude as the single particle spontaneous decay.  We handle this by transforming to Fourier space and working in the Lamb-Dicke (deep lattice) limit, which makes the dominant dipole-dipole term into a dispersion of collective atomic excitations.

We introduce an open system version of Feynman diagrams, and use them to calculate the steady state density matrix to lowest order in the Lamb-Dicke parameter.  We find that the steady state phonon number may be higher or lower than its non-interacting value, depending on the parameter settings, with a minimum of $\approx 30\%$ lower.

\appendix*
\section{Direct perturbative expansion} \label{App}

In this Appendix, we solve for the steady state density matrix by directly expanding the master equation in the Lamb-Dicke parameters $\eta_\alpha$, and verify that this gives the same result as earlier obtained using Feynman-like diagrams.

In the limit $\eta_\alpha\rightarrow 0$ (an infinitely deep lattice), we find that the atomic excitations are in a coherent state, while the phonons are stable free particles and hence can be in any state.  As we shall see below, to obtain a unique phonon steady state, we need to expand the master equation up to at least second order in $\eta_\alpha$.

In the following, we use the notation $\rho_\mathrm{ss}$ for the steady state density matrix and $\mathcal{L}$ for the time evolution superoperator, i.e.
\begin{eqnarray} \label{stationary}
\dot \rho_\mathrm{ss} \equiv \mathcal{L}(\rho_\mathrm{ss}) = 0 \, .
\end{eqnarray}
We then expand both in powers of $\eta_\alpha$,
\begin{eqnarray} 
\rho_\mathrm{ss} &=& \rho^{(0)}_\mathrm{ss}+\rho^{(1)}_\mathrm{ss}+\rho^{(2)}_\mathrm{ss} +\ldots \, , \notag \\
\mathcal{L} &=& \mathcal{L}^{(0)} + \mathcal{L}^{(1)}+ \mathcal{L}^{(2)}+\ldots \, .
\end{eqnarray}
As in the previous section, the superscripts indicate the order of the term with respect to $\eta_\alpha$.  This technique has previously been applied to single particle sideband cooling \cite{laser-cooling-theory,sw-sideband-theory}.  It is a form of adiabatic elimination \cite{laser-cooling-theory,sw-sideband-theory}, but does not neglect the spontaneous decay, which is important as cooling requires a decay process to carry away entropy.

\subsection{Zeroth order}

Substituting for $\mathcal{L}$ from the master equation (\ref{master}) and equating the zeroth order terms, we obtain
\begin{eqnarray} \label{master2}
- {\mathrm{i} \over \hbar} \big[ \, H_\mathrm{cond}^{(0)} \, \rho^{(0)}_\mathrm{ss} - \rho^{(0)}_\mathrm{ss} \, H_\mathrm{cond}^{(0) \dagger} \, \big] + \mathcal{R}^{(0)} (\rho^{(0)}_\mathrm{ss}) &=& 0 ~~~
\end{eqnarray}
with $H_\mathrm{cond}^{(0)} $ and $\mathcal{R}^{(0)}$ given in Eqs.~(\ref{Hconds}) and (\ref{Rs}). To solve this equation, we define 
\begin{eqnarray} \label{stilde}
\tilde{s}_{\mathbf{k}_\mathrm{L}} \equiv s_{\mathbf{k}_\mathrm{L}}+{\mathrm{i} \sqrt{N} \Omega \over c(\mathbf{k}_\mathrm{L})} ~~
\mathrm{and} ~~ \tilde{s}_\mathbf{p} \equiv s_\mathbf{p} ~ \forall ~ \mathbf{p}\neq\mathbf{k}_\mathrm{L} \, ,
\end{eqnarray}
and its Hermitian conjugate $\tilde{s}^\dagger_\mathbf{p}$.  This eliminates the $\Omega$ terms from $H_\mathrm{cond}^{(0)}$, giving
\begin{eqnarray}
H_\mathrm{cond}^{(0)} &=& - \sum\limits_\mathbf{p}\frac{\mathrm{i}}{2} \, \hbar c(\mathbf{p}) \, \tilde{s}^\dagger_\mathbf{p}\tilde{s}_\mathbf{p}+\sum\limits_{\mathbf{p},\alpha} \hbar \nu_\alpha \, b^\dagger_{\mathbf{p}\alpha}b_{\mathbf{p}\alpha} \, , \nonumber \\
\mathcal{R}^{(0)} (\rho) &=& \sum\limits_\mathbf{p}\mathrm{Re} \, c(\mathbf{p}) \, \tilde{s}_\mathbf{p}\rho \tilde{s}^\dagger_\mathbf{p} \, .
\end{eqnarray}
In terms of Feynman diagrams, this transformation is equivalent to cutting off the one-line (zeroth order laser) vertices and their associated lines, and absorbing their value (which is constant, as the lines always have wavevector $\mathbf{k}_\mathrm{L}$ and energy $0$) into the coefficient of the vertex they are attached to (Fig.~\ref{feynman4}(a)).

We now neglect the hardcore constraint of no more than one excitation per atom and treat the $\tilde{s}^\dagger_\mathbf{p}$ excitations as bosons. This is valid in the limit of low excitation densities \cite{Holstein,Shah}, or equivalently weak laser driving.  The zeroth order master equation (Eq.~(\ref{master2})) is then simply a sum over atom and phonon modes,
\begin{eqnarray}
\mathcal{L}^{(0)}&=&\sum\limits_\mathbf{p}\mathcal{L}^{(0)}_{\mathrm{at},\mathbf{p}}+\sum\limits_{\mathbf{p},\alpha}\mathcal{L}^{(0)}_{\mathrm{phn},\mathbf{p}\alpha} 
\end{eqnarray}
with the atom and phonon operators given by
\begin{eqnarray}
\mathcal{L}^{(0)}_{\mathrm{at},\mathbf{p}} &=& - \frac{1}{2} \,c(\mathbf{p}) \, \tilde{s}^\dagger_\mathbf{p}\tilde{s}_\mathbf{p}\rho- \frac{1}{2} \,c^*(\mathbf{p}) \,\rho \tilde{s}^\dagger_\mathbf{p}\tilde{s}_\mathbf{p} \notag \\
&& +\mathrm{Re} \, c(\mathbf{p}) \, \tilde{s}_\mathbf{p}\rho \tilde{s}^\dagger_\mathbf{p} \, ,\nonumber \\
\mathcal{L}^{(0)}_{\mathrm{phn},\mathbf{p}\alpha}&=&-\mathrm{i} \nu_\alpha \, (b^\dagger_{\mathbf{p}\alpha}b_{\mathbf{p}\alpha}\rho-\rho b^\dagger_{\mathbf{p}\alpha}b_{\mathbf{p}\alpha}) \, .
\end{eqnarray}
Since the operators $\mathcal{L}^{(0)}_{\mathrm{at},\mathbf{p}}$ and $\mathcal{L}^{(0)}_{\mathrm{phn},\mathbf{p}\alpha} $ all act on different modes, they all commute with each other.  We can hence choose a basis of simultaneous eigenstates of all these terms, which will also be eigenstates of the whole. Denoting the eigenstates of $\mathcal{L}^{(0)}$ by $\rho_\lambda$ with eigenvalues $\lambda$ such that $\mathcal{L}^{(0)}(\rho_\lambda)=\lambda\rho_\lambda$, these are of the form
\begin{eqnarray}
\rho_\lambda&=&\bigotimes\limits_\mathbf{p} \rho_{\lambda\mathrm{at},\mathbf{p}}\otimes\bigotimes\limits_{\mathbf{p},\alpha}\rho_{\lambda\mathrm{phn},\mathbf{p}\alpha} \, ,\notag\\
\lambda&=&\sum\limits_\mathbf{p}\lambda_{\mathrm{at},\mathbf{p}}+\sum\limits_{\mathbf{p},\alpha}\lambda_{\mathrm{phn},\mathbf{p}\alpha} 
\end{eqnarray}
with 
\begin{eqnarray}
\mathcal{L}^{(0)}_{\mathrm{at},\mathbf{p}} \, \rho_{\lambda\mathrm{at},\mathbf{p}}&=&\lambda_{\mathrm{at},\mathbf{p}} \, \rho_{\lambda\mathrm{at},\mathbf{p}} \, ,\notag\\
\mathcal{L}^{(0)}_{\mathrm{phn},\mathbf{p}\alpha} \, \rho_{\lambda\mathrm{phn},\mathbf{p}\alpha}&=&\lambda_{\mathrm{phn},\mathbf{p}\alpha} \, \rho_{\lambda\mathrm{phn},\mathbf{p}\alpha} \, .
\end{eqnarray}
The relevant atomic eigenstates $\rho_{\lambda\mathrm{at},\mathbf{p}}$ and their respective eigenvalues $\lambda_{\mathrm{at},\mathbf{p}}$ are
\begin{eqnarray}
&& \hspace*{-0.4cm} \rho_{\lambda\mathrm{at},\mathbf{p}}= |\tilde 0_\mathbf{p}\rangle\langle\tilde 0_\mathbf{p}| ~~ \mathrm{with} ~~ \lambda_{\mathrm{at},\mathbf{p}}=0 \, ,\notag\\
&& \hspace*{-0.4cm}  \rho_{\lambda\mathrm{at},\mathbf{p}}=|\tilde n_\mathbf{p}\rangle\langle\tilde 0_\mathbf{p}| ~~ \mathrm{with} ~~ \lambda_{\mathrm{at},\mathbf{p}}=-\frac{n}{2}c(\mathbf{p}) \, ,\notag\\
&& \hspace*{-0.4cm}  \rho_{\lambda\mathrm{at},\mathbf{p}}=|\tilde 0_\mathbf{p}\rangle\langle\tilde n_\mathbf{p}| ~~ \mathrm{with} ~~ \lambda_{\mathrm{at},\mathbf{p}}=-\frac{n}{2}c^*(\mathbf{p}) \, ,\\
&& \hspace*{-0.4cm}  \rho_{\lambda\mathrm{at},\mathbf{p}}=|\tilde 1_\mathbf{p}\rangle\langle\tilde 1_\mathbf{p}|-|\tilde 0_\mathbf{p}\rangle\langle\tilde 0_\mathbf{p}|
~~ \mathrm{with} ~~ \lambda_{\mathrm{at},\mathbf{p}}=-\mathrm{Re} \, c(\mathbf{p}) \, .\notag ~~~~
\end{eqnarray}
Here $|\tilde n_\mathbf{p}\rangle=\frac{1}{\sqrt{n!}}(\tilde{s}^\dagger_\mathbf{p})^n|\tilde 0_\mathbf{p}\rangle$ is the state with $n$ excitations in the mode $\mathbf{p}$. The tildes indicate that these are in terms of the $\tilde{s}_\mathbf{p}$ operators.  In terms of the original $s_\mathbf{p}$, $|\tilde 0_\mathbf{p} \rangle$ is a coherent state of amplitude $\sqrt{N}\Omega/|c(\mathbf{k}_\mathrm{L})|$ for $\mathbf{p}=\mathbf{k}_\mathrm{L}$, and a vacuum state for all other wavevectors $\mathbf{p}$.

Note that $\mathcal{L}^{(0)}$ does not correspond to a unitary evolution, and its eigenstates are hence not orthogonal.  In particular, if $\mathcal{P}$ is the projection onto the zeroth order steady states $\mathcal{L}^{(0)}=0$, $\mathcal{P}(|\tilde 1_\mathbf{p}\rangle\langle\tilde 1_\mathbf{p}|)=\mathcal{P}(|\tilde 1_\mathbf{p}\rangle\langle\tilde 1_\mathbf{p}|-|\tilde 0_\mathbf{p}\rangle\langle\tilde 0_\mathbf{p}|+|\tilde 0_\mathbf{p}\rangle\langle\tilde 0_\mathbf{p}|)=|\tilde 0_\mathbf{p}\rangle\langle\tilde 0_\mathbf{p}|$, not zero.

The phonon eigenstates are simply given by
\begin{eqnarray}
\rho_{\lambda \mathrm{phn},\mathbf{p}\alpha}=|m\rangle\langle n|
~~ \mathrm{with} ~~ \lambda_{\mathrm{phn},\mathbf{p}\alpha}=\mathrm{i} \nu_\alpha(n-m) \, . ~~~
\end{eqnarray}
Since the eigenvalue only depends on the difference in phonon energy between the right and left sides of $\rho$, and phonons of all wavevectors $\mathbf{p}$ have the same zeroth order energy $\nu_\alpha$, these phonon eigenstates are highly degenerate. In particular, the zeroth order steady states $\rho_\mathrm{ss}=\rho_{\lambda=0}$ are
\begin{eqnarray} \label{cmp}
\rho^{(0)}_\mathrm{ss} &=& \rho^{(0)}_\mathrm{phn} \otimes |\tilde 0 \rangle \langle \tilde 0| \, ,
\end{eqnarray}
where the atomic part $|\tilde 0 \rangle=\bigotimes\limits_\mathbf{p}|\tilde 0_\mathbf{p}\rangle$ is the vacuum state of all the $\tilde{s}_\mathbf{p}$, while the phonon part $\rho^{(0)}_\mathrm{phn}$ can be any phonon state which is diagonal in the total phonon energy (i.e. it can be a mixture, but not a superposition, of energies),
\begin{eqnarray} \label{cmp2}
\sum\limits_{\mathbf{p},\alpha} \hbar\nu_\alpha \, b^\dagger_{\mathbf{p}\alpha}b_{\mathbf{p}\alpha} \, \rho^{(0)}_\mathrm{phn} &=& \sum\limits_{\mathbf{p},\alpha} \hbar\nu_\alpha \, \rho^{(0)}_\mathrm{phn} \, b^\dagger_{\mathbf{p}\alpha}b_{\mathbf{p}\alpha} \, . ~~~
\end{eqnarray}
We see below that this degeneracy in the steady state is broken by the higher order terms, giving a unique $\rho^{(0)}_\mathrm{phn}$ when terms up to second order in $\eta_\alpha$ are considered.

\subsection{First order}

The first order terms in $\eta_\alpha$ in Eq.~(\ref{stationary}) are
\begin{eqnarray}
\mathcal{L}^{(0)} (\rho^{(1)}_\mathrm{ss})+\mathcal{L}^{(1)}(\rho^{(0)}_\mathrm{ss})=0 \, .
\end{eqnarray}
The unknown in this equation is $\rho^{(1)}_\mathrm{ss}$, and solving for it gives
\begin{eqnarray} \label{rho1ss}
\rho^{(1)}_\mathrm{ss}=- \left[ \mathcal{L}^{(0)} \right]^{-1}(\mathcal{L}^{(1)}(\rho^{(0)}_\mathrm{ss})) \, .
\end{eqnarray}
Here $\mathcal{L}^{(1)}$ is formed in the same way as $\mathcal{L}^{(0)}$ (in Eq.~(\ref{master2})), but using the first order terms $H_\mathrm{cond}^{(1)}$ and $\mathcal{R}^{(1)}(\rho)$, also given in Eqs.~(\ref{Hconds}) and (\ref{Rs}). The relevant terms, i.e. those which are non-zero when applied to $\rho^{(0)}_\mathrm{ss}$, are extracted in Eq.~(\ref{A4}).  Because $\mathcal{L}^{(0)}$ has zero eigenvalues, the inverse $\left[ \mathcal{L}^{(0)} \right]^{-1}$ is not defined on every density matrix.  However, the $\mathcal{L}^{(0)}=0$ states $\rho^{(0)}_\mathrm{ss}$ are diagonal in the total phonon energy (Eq.~(\ref{cmp2})). Every term in $\mathcal{L}^{(1)}$ either creates or destroys exactly one phonon. These terms hence transform $\mathcal{L}^{(0)}=0$ states to only $\mathcal{L}^{(0)}\neq 0$ states, on which the inverse $\left[ \mathcal{L}^{(0)} \right]^{-1}$ is defined. This means all the degenerate zeroth order steady states $\rho^{(0)}_\mathrm{ss}$ give valid first order steady states.  Hence, the first order terms do not break the degeneracy (non-uniqueness) of $\rho^{(0)}_\mathrm{phn}$.

\subsection{Second order}

We therefore consider the second order terms in Eq.~(\ref{stationary}), which we will find do break the degeneracy to give a unique steady state solution. These terms are given by 
\begin{equation}
0=\mathcal{L}^{(0)}(\rho^{(2)}_\mathrm{ss})+\mathcal{L}^{(1)}(\rho^{(1)}_\mathrm{ss})+\mathcal{L}^{(2)}(\rho^{(0)}_\mathrm{ss}) \, .
\end{equation}
Substituting $\rho^{(1)}_\mathrm{ss}$ from Eq.~(\ref{rho1ss}) and solving for $\rho^{(2)}_\mathrm{ss}$, we find that
\begin{equation}
\rho^{(2)}_\mathrm{ss} = - \left[ \mathcal{L}^{(0)} \right]^{-1} \left( \left\{\mathcal{L}^{(2)} -\mathcal{L}^{(1)} \left[ \mathcal{L}^{(0)} \right]^{-1} \mathcal{L}^{(1)} \right\} (\rho^{(0)}_\mathrm{ss}) \right) \, .\label{rho2ss}
\end{equation}
The $\mathcal{L}^{(2)}$ term comes from processes with a single second order phonon vertex (the first diagram in Fig.~\ref{feynman3}), while $\mathcal{L}^{(1)} \left[ \mathcal{L}^{(0)} \right]^{-1} \mathcal{L}^{(1)}$ comes from processes with two first order vertices.
Eq.~(\ref{rho2ss}) only has a solution if the inverse $[\mathcal{L}^{(0)}]^{-1}$ exists on the state it acts on, or equivalently if this state (in the large round brackets) has no component in the $\mathcal{L}^{(0)}=0$ subspace.  This condition can be written as
\begin{equation} \label{Pterms}
0 = \mathcal{P} \left( \left\{ \mathcal{L}^{(2)} -\mathcal{L}^{(1)} \left[ \mathcal{L}^{(0)} \right]^{-1} \mathcal{L}^{(1)} \right\} (\rho^{(0)}_\mathrm{ss} ) \right) \, ,
\end{equation}
where $\mathcal{P}$ is the (non-orthogonal) projection operator which selects all the density matrices that are of the form of $\rho^{(0)}_\mathrm{ss}$ in Eq.~(\ref{cmp}). Otherwise, the operator $[\mathcal{L}^{(0)}]^{-1}$ cannot be applied to this expression.

\subsection{The $\mathcal{P \, L}^{(2)}$ term in Eq.~(\ref{Pterms})}

In the following we summarise the terms which contribute to the evaluation of the first term in Eq.~(\ref{Pterms}). 
Only terms which are different from zero when applied to the vacuum state $|\tilde 0 \rangle$ of the atoms have to be taken into account, since $\mathcal{L}^{(2)}$ is applied to this state. Furthermore, the final state of the atoms should be $|\tilde 0 \rangle$, since the projector $\mathcal{P}$ selects terms which fulfill Eqs.~(\ref{cmp}). We also only need to consider phonon operators with an equal number of phonon creation and annihilation operators, as only these keep the total phonon energy diagonal, as described by Eq.~(\ref{cmp2}). Using Eqs.~(\ref{Hconds}) and (\ref{Rs}), we see that the relevant terms for the calculation of the $\mathcal{P \, L}^{(2)}$ term are given by 
\begin{widetext}
\begin{eqnarray} \label{A3}
H_\mathrm{cond}^{(2)} &=&\sum\limits_{\mathbf{p},\alpha,\beta} \frac{\mathrm{i}\hbar \Omega^2}{|c(\mathbf{k}_\mathrm{L})|^2} \left[ -\frac{\mathrm{i}}{2} \, \mathrm{Im} \, c(\mathbf{k}_\mathrm{L}) \, \eta_\alpha\eta_\beta\, \hat k_{\mathrm{L} \alpha}\hat k_{\mathrm{L} \beta} -e_{\alpha\beta}(\mathbf{k}_\mathrm{L})+e_{\alpha\beta}(\mathbf{k}_\mathrm{L} + \mathbf{p}) \right] (b_{\mathbf{p}\alpha} b^\dagger_{\mathbf{p} \beta} +b^\dagger_{-\mathbf{p} \alpha} b_{-\mathbf{p}\beta}) + \ldots \, , \notag \\
\mathcal{R}^{(2)}(\rho) &=& \sum\limits_{\mathbf{p},\alpha,\beta} \frac{\Omega^2}{|c(\mathbf{k}_\mathrm{L})|^2} \left[\mathrm{Re} \, e_{\alpha\beta}(\mathbf{k}_\mathrm{L})(b_{\mathbf{p}\alpha} b^\dagger_{\mathbf{p} \beta}
+b^\dagger_{-\mathbf{p} \alpha} b_{-\mathbf{p}\beta}) \, \rho -2\mathrm{Re} \, e_{\alpha\beta}(\mathbf{k}_\mathrm{L} + \mathbf{p})(b_{\mathbf{p}\alpha} \, \rho \, b^\dagger_{\mathbf{p} \beta} + b^\dagger_{-\mathbf{p} \alpha} \, \rho \, b_{-\mathbf{p}\beta}) \right. \notag \\
&&\left. +\mathrm{Re} \, e_{\alpha\beta}(\mathbf{k}_\mathrm{L})  \, \rho \, (b_{\mathbf{p}\alpha} b^\dagger_{\mathbf{p} \beta} +b^\dagger_{-\mathbf{p} \alpha} b_{-\mathbf{p}\beta})  \right] + \ldots \, , 
\end{eqnarray}
where the dots stand for terms which do not contribute to Eq.~(\ref{Pterms}), either because they are zero when acting on $\rho^{(0)}_\mathrm{ss}$, or because the projection $\mathcal{P}$ removes them.
Putting these together,
\begin{eqnarray}\label{PL2}
\mathcal{P L}^{(2)}(\rho^{(0)}_\mathrm{ss} ) &=&\sum\limits_{\mathbf{p},\alpha,\beta} \frac{\Omega^2}{|c(\mathbf{k}_\mathrm{L})|^2} \left[ -\frac{\mathrm{i}}{2} \, \mathrm{Im} \, c(\mathbf{k}_\mathrm{L}) \, \eta_\alpha\eta_\beta\, \hat k_{\mathrm{L} \alpha}\hat k_{\mathrm{L} \beta} -\mathrm{i} \, \mathrm{Im} \, e_{\alpha\beta}(\mathbf{k}_\mathrm{L})+e_{\alpha\beta}(\mathbf{k}_\mathrm{L} + \mathbf{p}) \right] (b_{\mathbf{p}\alpha} b^\dagger_{\mathbf{p} \beta} +b^\dagger_{-\mathbf{p} \alpha} b_{-\mathbf{p}\beta})\, \rho^{(0)}_\mathrm{ss}\notag \\
&&+\mathrm{H.c.}-\sum\limits_{\mathbf{p},\alpha,\beta} \frac{2\Omega^2}{|c(\mathbf{k}_\mathrm{L})|^2}\mathrm{Re} \, e_{\alpha\beta}(\mathbf{k}_\mathrm{L} + \mathbf{p})(b_{\mathbf{p}\alpha} \, \rho^{(0)}_\mathrm{ss} \, b^\dagger_{\mathbf{p} \beta} + b^\dagger_{-\mathbf{p} \alpha} \, \rho^{(0)}_\mathrm{ss} \, b_{-\mathbf{p}\beta}), 
\end{eqnarray}
These operators do not contain atomic operators, and are quadratic in phonon operators, with every product being diagonal in phonon wavevector $\mathbf{p}$.  In an isotropic lattice ($\nu_x=\nu_y=\nu_z$) $\alpha$ and $\beta$ can be different, but in a fully anisotropic lattice ($\nu_x$, $\nu_y$, $\nu_z$ all different) we must have $\alpha=\beta$ to stay diagonal in phonon energy (Eq.~(\ref{cmp2})).

\subsection{The $\mathcal{P \, L}^{(1)} \left[ \mathcal{L}^{(0)} \right]^{-1} \mathcal{L}^{(1)}$ term in Eq.~(\ref{Pterms})}

We next calculate $\mathcal{P} \left( \mathcal{L}^{(1)} \left[ \mathcal{L}^{(0)} \right]^{-1} \mathcal{L}^{(1)} (\rho^{(0)}_\mathrm{ss} ) \right)$ by applying these operators one at a time. For the first application of $\mathcal{L}^{(1)}$, again using Eqs.~(\ref{Hconds}) and (\ref{Rs}), we find
\begin{eqnarray} \label{A4}
H_\mathrm{cond}^{(1)} &=& \sum\limits_{\mathbf{p},\alpha} \frac{\hbar \Omega}{2} \, \left( \mathrm{i} \eta_\alpha \, \hat k_{\mathrm{L} \alpha} - {d_\alpha(\mathbf{k}_\mathrm{L})-d_\alpha(\mathbf{p}+\mathbf{k}_\mathrm{L}) \over c(\mathbf{k}_\mathrm{L})} \right) (b_{\mathbf{p}\alpha}+b^\dagger_{-\mathbf{p} \alpha}) s^\dagger_{\mathbf{p}+\mathbf{k}_\mathrm{L}}  - \sum\limits_{\alpha} \frac{\hbar \eta_\alpha \sqrt{N} \Omega^2 }{2 c(\mathbf{k}_\mathrm{L})} \, \hat k_{\mathrm{L} \alpha} (b_{\mathbf{0} \alpha}+b^\dagger_{{\mathbf{0}}\alpha}) +\ldots\, , \notag \\
\mathcal{R}^{(1)}(\rho) &=& \sum\limits_{\alpha}\frac{\mathrm{i} \sqrt{N} \Omega^2}{|c(\mathbf{k}_\mathrm{L})|^2} \, \mathrm{Im} \, d_\alpha(\mathbf{k}_\mathrm{L}) \, \left[  \rho \, (b_{\mathbf{0}\alpha}+b^\dagger_{\mathbf{0}\alpha})  - (b_{\mathbf{0}\alpha}+b^\dagger_{\mathbf{0}\alpha})\, \rho \right] +\ldots\, ,
\end{eqnarray}
where $\ldots$ stands for terms which give zero when applied to the vacuum state $|\tilde 0 \rangle$ of the atoms, and hence can be discarded.  As each term of $\mathcal{R}^{(1)}(\rho)$ in Eq.~(\ref{A4}) contains operators acting on either the left or the right of $\rho$, not both, these terms can be absorbed into the conditional Hamiltonian, giving
\begin{eqnarray} \label{A5}
\tilde H_\mathrm{cond}^{(1)} &=& \sum\limits_{\mathbf{p},\alpha} \frac{\hbar \Omega}{2} \, \left( \mathrm{i} \eta_\alpha \, \hat k_{\mathrm{L} \alpha} - {d_\alpha(\mathbf{k}_\mathrm{L})-d_\alpha(\mathbf{p}+\mathbf{k}_\mathrm{L}) \over c(\mathbf{k}_\mathrm{L})} \right) (b_{\mathbf{p}\alpha}+b^\dagger_{-\mathbf{p} \alpha}) \tilde s^\dagger_{\mathbf{p}+\mathbf{k}_\mathrm{L}} \notag \\
&& + \sum\limits_{\alpha} \frac{\hbar \sqrt{N} \Omega^2}{|c(\mathbf{k}_\mathrm{L})|^2} \left(  \mathrm{Im} \, d_\alpha(\mathbf{k}_\mathrm{L}) - \eta_\alpha  \, \mathrm{Re} \, c(\mathbf{k}_\mathrm{L}) \, \hat k_{\mathrm{L} \alpha} \right) (b_{\mathbf{0} \alpha}+b^\dagger_{{\mathbf{0}}\alpha})+\ldots
\end{eqnarray}
and $\mathcal{\tilde R}^{(1)}(\rho) = 0$.  The master equation is in both cases exactly the same.

Each of the four operator terms in $\tilde H_\mathrm{cond}^{(1)}$ takes $\rho^{(0)}_\mathrm{ss}$ to another eigenstate of $\mathcal{L}^{(0)}$; for example, $b_{\mathbf{p}\alpha}\tilde s^\dagger_{\mathbf{p}+\mathbf{k}_\mathrm{L}} \rho^{(0)}_\mathrm{ss}$ has the eigenvalue $-c(\mathbf{p}+\mathbf{k}_\mathrm{L})/2+\mathrm{i} \nu_\alpha$, the first term from $|\tilde 1\rangle\langle\tilde 0|=\tilde{s}^\dagger_{\mathbf{p}+\mathbf{k}_\mathrm{L}}|\tilde 0\rangle\langle\tilde 0|$ and the second term from the $b_{\mathbf{p}\alpha}$.  We can hence apply $\left[ \mathcal{L}^{(0)} \right]^{-1}$ by simply inserting the reciprocal eigenvalues. Using the short hand notation 
\begin{eqnarray}
\Xi &\equiv & \left[ \mathcal{L}^{(0)} \right]^{-1}  \left(  \tilde H_\mathrm{cond}^{(1)} \, \rho^{(0)}_\mathrm{ss} \right) \, ,
\end{eqnarray}
this yields
\begin{eqnarray}\label{LinvH}
\Xi &=& \left[ \sum\limits_{\mathbf{p},\alpha} \hbar \Omega \left( \mathrm{i} \eta_\alpha \, \hat k_{\mathrm{L} \alpha} - {d_\alpha(\mathbf{k}_\mathrm{L})-d_\alpha(\mathbf{p}+\mathbf{k}_\mathrm{L}) \over c(\mathbf{k}_\mathrm{L})} \right) \left(\frac{b_{\mathbf{p}\alpha}}{-c(\mathbf{p}+\mathbf{k}_\mathrm{L})+2\mathrm{i} \nu_\alpha}+\frac{b^\dagger_{-\mathbf{p} \alpha}}{-c(\mathbf{p}+\mathbf{k}_\mathrm{L})-2\mathrm{i} \nu_\alpha}\right) \tilde s^\dagger_{\mathbf{p}+\mathbf{k}_\mathrm{L}} \right. \notag \\
&& \left. - \sum\limits_{\alpha} \frac{\mathrm{i}\hbar \sqrt{N} \Omega^2}{\nu_\alpha |c(\mathbf{k}_\mathrm{L})|^2} \left(  \mathrm{Im} \, d_\alpha(\mathbf{k}_\mathrm{L}) - \eta_\alpha  \, \mathrm{Re} \, c(\mathbf{k}_\mathrm{L}) \, \hat k_{\mathrm{L} \alpha} \right) \left(b_{\mathbf{0} \alpha} - b^\dagger_{{\mathbf{0}}\alpha}\right) \right] \, \rho^{(0)}_\mathrm{ss} \, .
\end{eqnarray}
This inverse is not unique: we could add in an arbitrary $\mathcal{L}^{(0)}=0$ state.  However, doing so would have no effect on the final result, as it would be turned into an $\mathcal{L}^{(0)}\neq 0$ state by the second $\mathcal{L}^{(1)}$ then annihilated by $\mathcal{P}$.

We now apply the second $\mathcal{L}^{(1)}$, and finally apply $\mathcal{P}$, i.e. keep only terms which satisfy Eqs.~(\ref{cmp}) and (\ref{cmp2}). This gives 
\begin{eqnarray}
\mathcal{P} \left( \mathcal{L}^{(1)} \left[ \mathcal{L}^{(0)} \right]^{-1} \mathcal{L}^{(1)} (\rho^{(0)}_\mathrm{ss} ) \right)
&=& \mathcal{P} \left[ - {1 \over \hbar^2} \,  H_\mathrm{cond}^{(1)}\Xi+ {1 \over \hbar^2} \, \Xi H_\mathrm{cond}^{(1) \dagger} - {\mathrm{i} \over \hbar} \, \mathcal{R}^{(1)} \left( \Xi\right) \right] + \mathrm{H.c.} \, ,
\end{eqnarray}
where $\Xi $ is defined by Eq.~(\ref{LinvH}) and the Hermitian conjugate comes from $ \left[ \mathcal{L}^{(0)} \right]^{-1}  \left(  \rho^{(0)}_\mathrm{ss}\tilde H_\mathrm{cond}^{(1)\dagger} \right)=\Xi^\dagger $. Using Eq.~(\ref{Hconds}), we find that the first term in this equation is given by
\begin{eqnarray}
\mathcal{P} \left[  \tilde H_\mathrm{cond}^{(1)} \Xi \right] &=&  \left[ \sum\limits_{\mathbf{p},\alpha,\beta} \frac{\hbar^2 \Omega^2}{2}  \, \left(-\mathrm{i}\eta_\beta \, \hat k_{\mathrm{L} \beta} +\frac{d_\beta(\mathbf{k}_\mathrm{L}+\mathbf{p})-d_\beta(\mathbf{k}_\mathrm{L})}{c^*(\mathbf{k}_\mathrm{L})}\right)\left( \mathrm{i} \eta_\alpha \, \hat k_{\mathrm{L} \alpha} - {d_\alpha(\mathbf{k}_\mathrm{L})-d_\alpha(\mathbf{p}+\mathbf{k}_\mathrm{L}) \over c(\mathbf{k}_\mathrm{L})}  \right)  \right. \notag\\
&&\times \left(\frac{b^\dagger_{\mathbf{p} \beta}b_{\mathbf{p}\alpha}}{-c(\mathbf{p}+\mathbf{k}_\mathrm{L})+2\mathrm{i} \nu_\alpha}+\frac{b_{\mathbf{-p}\beta}b^\dagger_{-\mathbf{p} \alpha}}{-c(\mathbf{p}+\mathbf{k}_\mathrm{L})-2\mathrm{i} \nu_\alpha}\right) + \sum\limits_{\alpha,\beta} \frac{\mathrm{i} \hbar^2 N \Omega^4}{\nu_\alpha |c(\mathbf{k}_\mathrm{L})|^4} \eta_\beta  \, \mathrm{Re} \, c(\mathbf{k}_\mathrm{L}) \, \hat k_{\mathrm{L} \beta} \notag \\
&& \left. \times \left(  \mathrm{Im} \, d_\alpha(\mathbf{k}_\mathrm{L}) - \eta_\alpha  \, \mathrm{Re} \, c(\mathbf{k}_\mathrm{L}) \, \hat k_{\mathrm{L} \alpha} \right) \left(b^\dagger_{{\mathbf{0}}\beta}b_{\mathbf{0} \alpha} - b_{\mathbf{0} \beta}b^\dagger_{{\mathbf{0}}\alpha}\right) \right] \, \rho^{(0)}_\mathrm{ss} \, . ~~~
\end{eqnarray}
Analogously, remembering that $\mathcal{P}(|\tilde 1\rangle\langle\tilde 1|)=|\tilde 0\rangle\langle\tilde 0|$, we find that 
\begin{eqnarray}
\mathcal{P} \left[ \Xi \tilde H_\mathrm{cond}^{(1) \dagger} \right] &=& \sum\limits_{\mathbf{p},\alpha,\beta} \frac{\hbar^2 \Omega^2}{2}  \, \left(-\mathrm{i}\eta_\beta \, \hat k_{\mathrm{L} \beta} +\frac{d^*_\beta(\mathbf{k}_\mathrm{L}+\mathbf{p})-d^*_\beta(\mathbf{k}_\mathrm{L})}{c^*(\mathbf{k}_\mathrm{L})}\right) \left( \mathrm{i} \eta_\alpha \, \hat k_{\mathrm{L} \alpha} - {d_\alpha(\mathbf{k}_\mathrm{L})-d_\alpha(\mathbf{p}+\mathbf{k}_\mathrm{L}) \over c(\mathbf{k}_\mathrm{L})} \right)\notag\\
&& \left(\frac{b_{\mathbf{p}\alpha}\, \rho^{(0)}_\mathrm{ss} \,b^\dagger_{\mathbf{p} \beta}}{-c(\mathbf{p}+\mathbf{k}_\mathrm{L})+2\mathrm{i} \nu_\alpha}+\frac{b^\dagger_{-\mathbf{p} \alpha}\, \rho^{(0)}_\mathrm{ss} \,b_{-\mathbf{p} \beta}}{-c(\mathbf{p}+\mathbf{k}_\mathrm{L})-2\mathrm{i} \nu_\alpha}\right) +  \sum\limits_{\alpha,\beta} \frac{\mathrm{i} \hbar^2 N \Omega^4}{\nu_\alpha |c(\mathbf{k}_\mathrm{L})|^4} \eta_\beta  \, \mathrm{Re} \, c(\mathbf{k}_\mathrm{L}) \, \hat k_{\mathrm{L} \beta} \notag \\
&& \times \left(  \mathrm{Im} \, d_\alpha(\mathbf{k}_\mathrm{L}) - \eta_\alpha  \, \mathrm{Re} \, c(\mathbf{k}_\mathrm{L}) \, \hat k_{\mathrm{L} \alpha} \right) \left(b_{\mathbf{0} \alpha}\, \rho^{(0)}_\mathrm{ss}\,b^\dagger_{{\mathbf{0}}\beta}- b^\dagger_{{\mathbf{0}}\alpha}  \, \rho^{(0)}_\mathrm{ss}\, b_{\mathbf{0} \beta} \right) . ~~~~~
\end{eqnarray}  
Moreover, we obtain from Eq.~(\ref{Rs}), 
\begin{eqnarray}
\mathcal{P} \mathcal{R}^{(1)} (\Xi) &=& -\sum\limits_{\mathbf{p},\alpha,\beta}\frac{\hbar \Omega^2}{c^*(\mathbf{k}_\mathrm{L})} \left( \mathrm{i} \eta_\alpha \, \hat k_{\mathrm{L} \alpha} - {d_\alpha(\mathbf{k}_\mathrm{L})-d_\alpha(\mathbf{p}+\mathbf{k}_\mathrm{L}) \over c(\mathbf{k}_\mathrm{L})}  \right) 
\left[ \mathrm{Im} \, d_\beta(\mathbf{p}+\mathbf{k}_\mathrm{L}) \,\left(\frac{b_{\mathbf{p}\alpha}\rho^{(0)}_\mathrm{ss} b^\dagger_{\mathbf{p}\beta}}{-c(\mathbf{p}+\mathbf{k}_\mathrm{L})+2\mathrm{i} \nu_\alpha} \right. \right. \notag\\
&& \left. \left. +\frac{b^\dagger_{-\mathbf{p} \alpha}\rho^{(0)}_\mathrm{ss} b_{-\mathbf{p}\beta}}{-c(\mathbf{p}+\mathbf{k}_\mathrm{L})-2\mathrm{i} \nu_\alpha} \right) - \mathrm{Im} \, d_\beta(\mathbf{k}_\mathrm{L}) \, \left(\frac{b^\dagger_{\mathbf{p}\beta}b_{\mathbf{p}\alpha}}{-c(\mathbf{p}+\mathbf{k}_\mathrm{L})+2\mathrm{i} \nu_\alpha}+\frac{b_{-\mathbf{p}\beta}b^\dagger_{-\mathbf{p} \alpha}}{-c(\mathbf{p}+\mathbf{k}_\mathrm{L})-2\mathrm{i} \nu_\alpha}\right)\rho^{(0)}_\mathrm{ss}\right] \notag \\
&&+\sum\limits_{\alpha,\beta} \frac{\hbar N \Omega^4}{\nu_\alpha |c(\mathbf{k}_\mathrm{L})|^4} \, \mathrm{Im} \, d_\beta(\mathbf{k}_\mathrm{L})  \left(  \mathrm{Im} \, d_\alpha(\mathbf{k}_\mathrm{L}) - \eta_\alpha  \, \mathrm{Re} \, c(\mathbf{k}_\mathrm{L}) \, \hat k_{\mathrm{L} \alpha} \right) \left( b_{\mathbf{0} \alpha}\rho^{(0)}_\mathrm{ss} \,b^\dagger_{\mathbf{0}\beta} -b^\dagger_{\mathbf{0}\beta}b_{\mathbf{0} \alpha}\,\rho^{(0)}_\mathrm{ss} - b^\dagger_{\mathbf{0}\alpha}\, \rho^{(0)}_\mathrm{ss} \,b_{\mathbf{0}\beta} \right. \notag\\
&& \left. +b_{\mathbf{0}\beta}b^\dagger_{\mathbf{0}\alpha} \, \rho^{(0)}_\mathrm{ss} \right).
\end{eqnarray}  
Putting these three equations together, we finally obtain
\begin{eqnarray} \label{PL101}
\mathcal{P} \left( \mathcal{L}^{(1)} \left[ \mathcal{L}^{(0)} \right]^{-1} \mathcal{L}^{(1)} (\rho^{(0)}_\mathrm{ss} ) \right)
&=& \sum\limits_{\mathbf{p},\alpha,\beta} \frac{\Omega^2}{2}  \, \left(\mathrm{i}\eta_\beta \, \hat k_{\mathrm{L} \beta} -\frac{d_\beta(\mathbf{k}_\mathrm{L}+\mathbf{p})-d_\beta^*(\mathbf{k}_\mathrm{L})}{c^*(\mathbf{k}_\mathrm{L})}\right)\left( \mathrm{i} \eta_\alpha \, \hat k_{\mathrm{L} \alpha} - {d_\alpha(\mathbf{k}_\mathrm{L})-d_\alpha(\mathbf{p}+\mathbf{k}_\mathrm{L}) \over c(\mathbf{k}_\mathrm{L})} \right)  \notag\\
&& \times \left(\frac{b^\dagger_{\mathbf{p} \beta}b_{\mathbf{p}\alpha}\rho^{(0)}_\mathrm{ss}-b_{\mathbf{p}\alpha}\, \rho^{(0)}_\mathrm{ss} \,b^\dagger_{\mathbf{p} \beta}}{-c(\mathbf{p}+\mathbf{k}_\mathrm{L})+2\mathrm{i} \nu_\alpha}+\frac{b_{\mathbf{-p}\beta}b^\dagger_{-\mathbf{p} \alpha}\, \rho^{(0)}_\mathrm{ss}-b^\dagger_{-\mathbf{p} \alpha}\, \rho^{(0)}_\mathrm{ss} \,b_{-\mathbf{p} \beta}}{-c(\mathbf{p}+\mathbf{k}_\mathrm{L})-2\mathrm{i} \nu_\alpha}\right) \notag \\
&&+\sum\limits_{\alpha,\beta}\frac{\mathrm{i}N\Omega^4}{\nu_\alpha|c(\mathbf{k}_\mathrm{L})|^4}\left(\mathrm{Im} \, d_\beta(\mathbf{k}_\mathrm{L})-\eta_\beta  \, \mathrm{Re} \, c(\mathbf{k}_\mathrm{L}) \, \hat k_{\mathrm{L} \beta}\right)  \left(  \mathrm{Im} \, d_\alpha(\mathbf{k}_\mathrm{L}) - \eta_\alpha  \, \mathrm{Re} \, c(\mathbf{k}_\mathrm{L}) \, \hat k_{\mathrm{L} \alpha} \right) \notag\\
&&~~\times\left(-b_{\mathbf{0} \alpha}\rho^{(0)}_\mathrm{ss} \,b^\dagger_{\mathbf{0}\beta}+b^\dagger_{\mathbf{0}\beta}b_{\mathbf{0} \alpha}\,\rho^{(0)}_\mathrm{ss}+b^\dagger_{\mathbf{0}\alpha}\, \rho^{(0)}_\mathrm{ss} \,b_{\mathbf{0}\beta}-b_{\mathbf{0}\beta}b^\dagger_{\mathbf{0}\alpha} \, \rho^{(0)}_\mathrm{ss} \, \right) +\text{H.c.}
\end{eqnarray}
Again, $\alpha$ and $\beta$ are only independent in the isotropic case; in the fully anisotropic case we must have $\alpha=\beta$.

\subsection{The heating and cooling rates of the different phonon modes}

Substituting Eqs.~(\ref{PL2}) and (\ref{PL101}) into Eq.~(\ref{Pterms}), and collecting terms acting on the same phonon mode by relabelling $-\mathbf{p}\rightarrow \mathbf{p}$ where necessary, we obtain for the phonon steady state
\begin{eqnarray} \label{final}
0&=&\frac{1}{2}\sum\limits_{\mathbf{p},\alpha,\beta}\left[M_{\mathrm{h} \alpha\beta}(\mathbf{p})(b^\dagger_{\mathbf{p} \alpha}\, \rho^{(0)}_\mathrm{ss} \,b_{\mathbf{p} \beta}-b_{\mathbf{p}\beta}b^\dagger_{\mathbf{p} \alpha}\, \rho^{(0)}_\mathrm{ss})+M_{\mathrm{h} \alpha\beta}^*(\mathbf{p})(b^\dagger_{\mathbf{p} \beta}\, \rho^{(0)}_\mathrm{ss} \,b_{\mathbf{p} \alpha}-\rho^{(0)}_\mathrm{ss}\, b_{\mathbf{p}\alpha}b^\dagger_{\mathbf{p} \beta})\right.\notag\\
&&\left.+M_{\mathrm{c} \alpha\beta}(\mathbf{p})(b_{\mathbf{p} \alpha}\, \rho^{(0)}_\mathrm{ss} \,b^\dagger_{\mathbf{p} \beta}-b^\dagger_{\mathbf{p}\beta}b_{\mathbf{p} \alpha}\, \rho^{(0)}_\mathrm{ss})+M_{\mathrm{c} \alpha\beta}^*(\mathbf{p})(b_{\mathbf{p} \beta}\, \rho^{(0)}_\mathrm{ss} \,b^\dagger_{\mathbf{p} \alpha}-\rho^{(0)}_\mathrm{ss}\, b^\dagger_{\mathbf{p}\alpha}b_{\mathbf{p} \beta})\right] \, .
\end{eqnarray}
As this does not contain atomic operators, we can cancel the atomic part $|\tilde 0 \rangle \langle \tilde 0|$, which replaces $\rho^{(0)}_\mathrm{ss}$ by $\rho^{(0)}_\mathrm{phn}$.  It is also quadratic in phonon operators and diagonal in phonon wavevector.

For $\mathbf{p}\neq\mathbf{0}$, the coefficients $M_{\mathrm{h} \alpha\beta}(\mathbf{p})$ and $M_{\mathrm{c} \alpha\beta}(\mathbf{p})$ are given by
\begin{eqnarray} \label{M1}
M_{\mathrm{h} \alpha\beta}(\mathbf{p})&=&- \frac{\Omega^2}{c(\mathbf{k}_\mathrm{L}-\mathbf{p})+2\mathrm{i} \nu_\alpha}  \, \left(\mathrm{i}\eta_\beta \, \hat k_{\mathrm{L} \beta} -\frac{d_\beta(\mathbf{k}_\mathrm{L}-\mathbf{p})-d_\beta^*(\mathbf{k}_\mathrm{L})}{c^*(\mathbf{k}_\mathrm{L})}\right)\left( \mathrm{i} \eta_\alpha \, \hat k_{\mathrm{L} \alpha} - {d_\alpha(\mathbf{k}_\mathrm{L})-d_\alpha(\mathbf{k}_\mathrm{L}-\mathbf{p}) \over c(\mathbf{k}_\mathrm{L})} \right)\notag\\
&&-\frac{\Omega^2}{|c(\mathbf{k}_\mathrm{L})|^2} [-\mathrm{i}\mathrm{Im} \, c(\mathbf{k}_\mathrm{L}) \, \eta_\alpha\eta_\beta\, \hat k_{\mathrm{L} \alpha}\hat k_{\mathrm{L} \beta} -2\mathrm{i}\mathrm{Im} \, e_{\alpha\beta}(\mathbf{k}_\mathrm{L})+2e_{\alpha\beta}(\mathbf{k}_\mathrm{L} - \mathbf{p})]\notag,\\
M_{\mathrm{c} \alpha\beta}(\mathbf{p})&=&- \frac{\Omega^2}{c(\mathbf{k}_\mathrm{L}+\mathbf{p})-2\mathrm{i} \nu_\alpha}  \, \left(\mathrm{i}\eta_\beta \, \hat k_{\mathrm{L} \beta} -\frac{d_\beta(\mathbf{k}_\mathrm{L}+\mathbf{p})-d_\beta^*(\mathbf{k}_\mathrm{L})}{c^*(\mathbf{k}_\mathrm{L})}\right)\left( \mathrm{i} \eta_\alpha \, \hat k_{\mathrm{L} \alpha} - {d_\alpha(\mathbf{k}_\mathrm{L})-d_\alpha(\mathbf{k}_\mathrm{L}+\mathbf{p}) \over c(\mathbf{k}_\mathrm{L})} \right)\notag\\
&&-\frac{\Omega^2}{|c(\mathbf{k}_\mathrm{L})|^2} [-\mathrm{i}\mathrm{Im} \, c(\mathbf{k}_\mathrm{L}) \, \eta_\alpha\eta_\beta\, \hat k_{\mathrm{L} \alpha}\hat k_{\mathrm{L} \beta} -2\mathrm{i}\mathrm{Im} \, e_{\alpha\beta}(\mathbf{k}_\mathrm{L})+2e_{\alpha\beta}(\mathbf{k}_\mathrm{L} + \mathbf{p})] \, .
\end{eqnarray}
For $\mathbf{p} = \mathbf{0}$, there appear to be additional terms
\begin{eqnarray} \label{M2}
M_{\mathrm{h} \alpha\beta}(\mathbf{0})&=&  \frac{\eta_\alpha \, \hat k_{\mathrm{L} \alpha} \Omega^2}{c(\mathbf{k}_\mathrm{L})+2\mathrm{i} \nu_\alpha}  \, \left(\eta_\beta \, \hat k_{\mathrm{L} \beta} -\frac{2 \mathrm{Im} \, d_\beta(\mathbf{k}_\mathrm{L})}{c^*(\mathbf{k}_\mathrm{L})}\right) -\frac{\Omega^2}{|c(\mathbf{k}_\mathrm{L})|^2} [-\mathrm{i} \, \mathrm{Im} \, c(\mathbf{k}_\mathrm{L}) \, \eta_\alpha\eta_\beta\, \hat k_{\mathrm{L} \alpha}\hat k_{\mathrm{L} \beta} +2\mathrm{Re} \, e_{\alpha\beta}(\mathbf{k}_\mathrm{L})] \notag \\
&& - \frac{\mathrm{i}N\Omega^4}{\nu_\alpha|c(\mathbf{k}_\mathrm{L})|^4}\left(\mathrm{Im} \, d_\beta(\mathbf{k}_\mathrm{L})-\eta_\beta  \, \mathrm{Re} \, c(\mathbf{k}_\mathrm{L}) \, \hat k_{\mathrm{L} \beta}\right)  \left(  \mathrm{Im} \, d_\alpha(\mathbf{k}_\mathrm{L}) - \eta_\alpha  \, \mathrm{Re} \, c(\mathbf{k}_\mathrm{L}) \, \hat k_{\mathrm{L} \alpha} \right) \, , \notag \\
M_{\mathrm{c} \alpha\beta}(\mathbf{0})&=&  \frac{\eta_\alpha \, \hat k_{\mathrm{L} \alpha} \Omega^2}{c(\mathbf{k}_\mathrm{L})- 2\mathrm{i} \nu_\alpha}  \, \left(\eta_\beta \, \hat k_{\mathrm{L} \beta} -\frac{2 \mathrm{Im} \, d_\beta(\mathbf{k}_\mathrm{L})}{c^*(\mathbf{k}_\mathrm{L})}\right) -\frac{\Omega^2}{|c(\mathbf{k}_\mathrm{L})|^2} [-\mathrm{i} \, \mathrm{Im} \, c(\mathbf{k}_\mathrm{L}) \, \eta_\alpha\eta_\beta\, \hat k_{\mathrm{L} \alpha}\hat k_{\mathrm{L} \beta} +2\mathrm{Re} \, e_{\alpha\beta}(\mathbf{k}_\mathrm{L})] \notag \\
&& + \frac{\mathrm{i}N\Omega^4}{\nu_\alpha|c(\mathbf{k}_\mathrm{L})|^4}\left(\mathrm{Im} \, d_\beta(\mathbf{k}_\mathrm{L})-\eta_\beta  \, \mathrm{Re} \, c(\mathbf{k}_\mathrm{L}) \, \hat k_{\mathrm{L} \beta}\right)  \left(  \mathrm{Im} \, d_\alpha(\mathbf{k}_\mathrm{L}) - \eta_\alpha  \, \mathrm{Re} \, c(\mathbf{k}_\mathrm{L}) \, \hat k_{\mathrm{L} \alpha} \right)\, .
\end{eqnarray}
However, these extra terms (the last line of each equation) are pure imaginary, symmetric in $\alpha,\beta$ and of opposite signs in $M_{\mathrm{h} \alpha\beta}$ and $M_{\mathrm{c} \alpha\beta}$.  As $b_{\mathbf{p}\alpha}b^\dagger_{\mathbf{p} \beta}-b^\dagger_{\mathbf{p}\beta}b_{\mathbf{p} \alpha}=\delta_{\alpha\beta}$ is a number so commutes with $\rho$, such terms cancel out in Eq.~(\ref{final}), so Eq.~(\ref{M1}) can also be used for $\mathbf{p} = \mathbf{0}$.

In the fully anisotropic case (three different phonon frequencies $\nu_x$, $\nu_y$, $\nu_z$), Eq.~(\ref{final}) simplifies to
\begin{eqnarray} \label{43}
0&=&\sum\limits_{\mathbf{p},\alpha}\left[-\frac{1}{2}M_{\mathrm{h} \alpha\alpha}(\mathbf{p})b_{\mathbf{p}\alpha}b^\dagger_{\mathbf{p} \alpha}\, \rho^{(0)}_\mathrm{ss} -\frac{1}{2}M_{\mathrm{h} \alpha\alpha}^*(\mathbf{p})\rho^{(0)}_\mathrm{ss}\, b_{\mathbf{p}\alpha}b^\dagger_{\mathbf{p} \alpha} + \mathrm{Re} \, M_{\mathrm{h} \alpha\alpha}(\mathbf{p})b^\dagger_{\mathbf{p} \alpha}\, \rho^{(0)}_\mathrm{ss} \,b_{\mathbf{p} \alpha}\right.  \notag\\
&& \left. -\frac{1}{2}M_{\mathrm{c} \alpha\alpha} (\mathbf{p}) b^\dagger_{\mathbf{p}\alpha}b_{\mathbf{p} \alpha}\, \rho^{(0)}_\mathrm{ss}-\frac{1}{2}M_{\mathrm{c} \alpha\alpha}^*(\mathbf{p})\rho^{(0)}_\mathrm{ss}\, b^\dagger_{\mathbf{p}\alpha}b_{\mathbf{p} \alpha}  +\mathrm{Re} \, M_{\mathrm{c} \alpha\alpha}(\mathbf{p})b_{\mathbf{p} \alpha}\, \rho^{(0)}_\mathrm{ss} \,b^\dagger_{\mathbf{p} \alpha} \right] \, .
\end{eqnarray}
In this case each phonon mode $\mathbf{p} \alpha$ evolves independently, with a heating (phonon creation) rate $\mathrm{Re} \, M_{\mathrm{h} \alpha\alpha} (\mathbf{p})$, a cooling (phonon decay) rate $\mathrm{Re} \, M_{\mathrm{c} \alpha\alpha} (\mathbf{p})$, and an energy dispersion $\mathrm{Im} \, M_{\mathrm{h} \alpha\alpha} (\mathbf{p})+\mathrm{Im} \, M_{\mathrm{c} \alpha\alpha} (\mathbf{p})$.
\subsection{Consistency with diagrammatic method}
\begin{figure*}[t]
\includegraphics{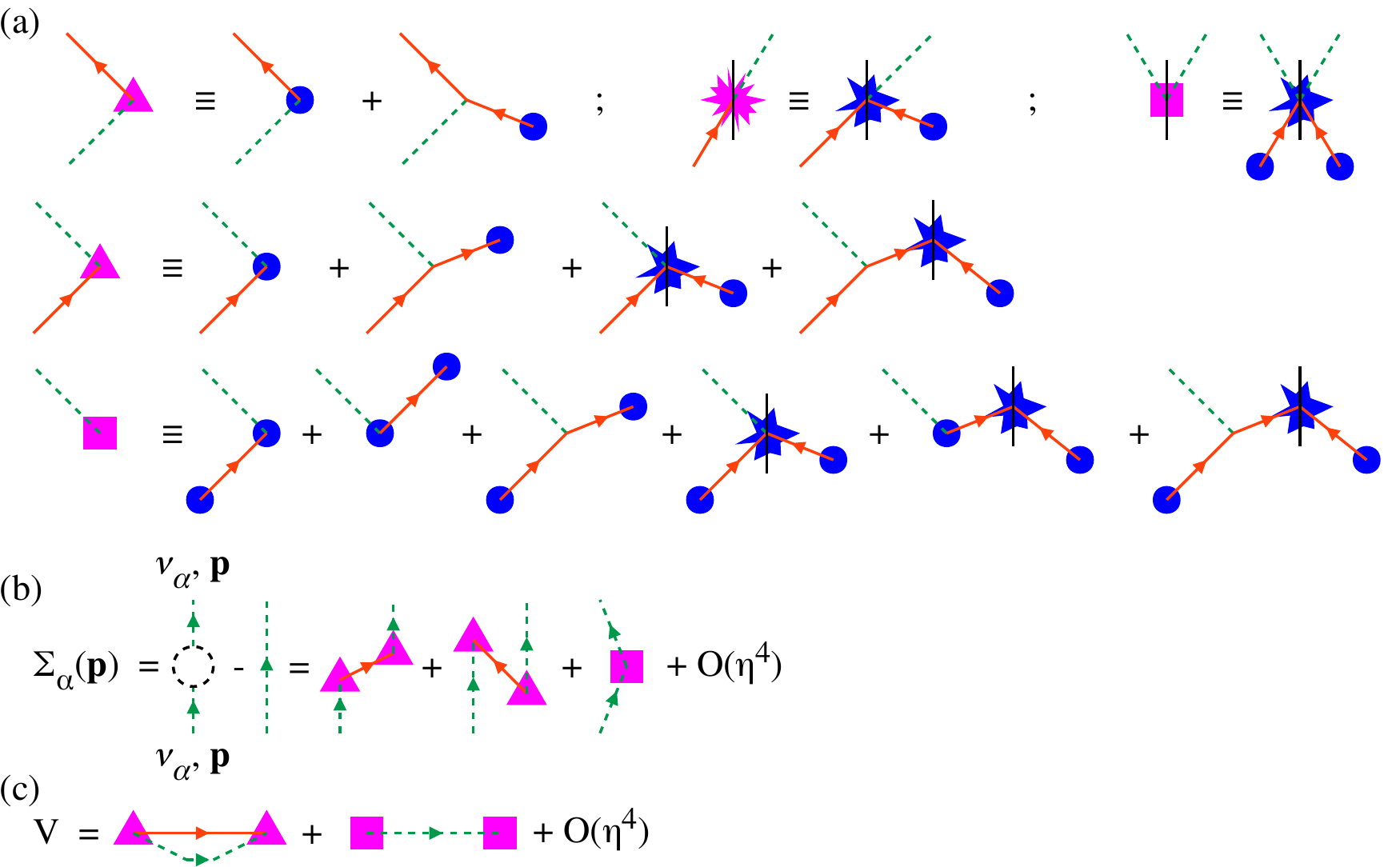}
\caption{(Color online) (a) Feynman diagram transformations corresponding to $s_\mathbf{p}\rightarrow \tilde s_\mathbf{p}$: the laser vertices are replaced by the vertices introduced here, the one-line (zeroth order laser) vertex being removed completely, while the hopping and decay vertices are unchanged.  All except the one at the top right have both 1-phonon and 2-phonon versions, with the latter having both phonon lines attached to the same vertex.  The reset vertices (the 2 in the top right, with $\rho$ lines through them) always have at least one line on each side of $\rho$, and hence the phonon-only reset vertex (top right) has no 1-phonon version.  (b) Diagrams for the phonon self-energy $\Sigma_\alpha(\mathbf{p})$.  (c) Diagrams for the vacuum bubble $V$.  The diagrams for $\Sigma_\alpha(\mathbf{p})$ are all on the left of the $\rho$ line (the same diagrams on the right give $\Sigma^*_\alpha(\mathbf{p})$), while those for $V$ are summed over both sides of $\rho$.  As in Fig.~\ref{feynman}, solid orange lines are atomic excitations, dotted green lines are phonons (with either direction allowed if there is no arrow), blue circles are laser driving, blue stars are spontaneous decay. \label{feynman4}}
\end{figure*}
Comparing Eq.~(\ref{M1}) with Eq.~(\ref{feynmansum}), we see that both methods give the same heating and cooling rates, i.e. $\mathrm{Re} \, M_{\mathrm{h} \alpha\alpha}=R_{\mathrm{h}\alpha},\,\mathrm{Re} \, M_{\mathrm{c} \alpha\alpha}=R_{\mathrm{c}\alpha}$.

The self-energy correction to the phonon propagation line, as given by the diagrams in Fig.~\ref{feynman4}(b), is
\begin{eqnarray}
\Sigma_\alpha(\mathbf{p})&=&\frac{\Omega^2}{c(\mathbf{k}_\mathrm{L}+\mathbf{p})-2\mathrm{i}\nu_\alpha}\left(\eta_\alpha \hat k_{\mathrm{L}\alpha}+\mathrm{i}\frac{d_\alpha(\mathbf{k}_\mathrm{L})-d_\alpha(\mathbf{k}_\mathrm{L}+\mathbf{p})}{c(\mathbf{k}_\mathrm{L})}\right)\notag\\
&&\times\left(-\frac{1}{2}\eta_\alpha \hat k_{\mathrm{L}\alpha}-\mathrm{i}\frac{d_\alpha(\mathbf{k}_\mathrm{L})-d_\alpha(\mathbf{k}_\mathrm{L}+\mathbf{p})}{2c(\mathbf{k}_\mathrm{L})}+\frac{\mathrm{Im}\,d_\alpha(\mathbf{k}_\mathrm{L})}{c^*(\mathbf{k}_\mathrm{L})}+\mathrm{i}\mathrm{Re}\,c(\mathbf{k}_\mathrm{L})\frac{d_\alpha(\mathbf{k}_\mathrm{L})-d_\alpha(\mathbf{k}_\mathrm{L}+\mathbf{p})}{|c(\mathbf{k}_\mathrm{L})|^2}\right)\notag\\
&&+(\text{same with }\mathbf{p}\rightarrow -\mathbf{p},\, \nu_\alpha\rightarrow -\nu_\alpha)+\frac{\Omega^2}{c(\mathbf{k}_\mathrm{L})}\left\{\eta_\alpha^2\hat k_{\mathrm{L}\alpha}^2\left(1-\frac{\mathrm{Re}\,c(\mathbf{k}_\mathrm{L})}{c^*(\mathbf{k}_\mathrm{L})}\right)+\frac{2\mathrm{Re}\,e_{\alpha\alpha}(\mathbf{k}_\mathrm{L})}{c^*(\mathbf{k}_\mathrm{L})}\right.\notag\\
&&\left.+\left[2e_{\alpha\alpha}(\mathbf{k}_\mathrm{L})-e_{\alpha\alpha}(\mathbf{k}_\mathrm{L}-\mathbf{p})-e_{\alpha\alpha}(\mathbf{k}_\mathrm{L}+\mathbf{p})\right]\left(\frac{1}{c(\mathbf{k}_\mathrm{L})}-\frac{2\mathrm{Re}\,c(\mathbf{k}_\mathrm{L})}{|c(\mathbf{k}_\mathrm{L})|^2}\right)\right\}+O(\eta^4)\notag\\
&=&-\frac{1}{2}(M_{\mathrm{h} \alpha\alpha}(\mathbf{p})+M_{\mathrm{c} \alpha\alpha}(\mathbf{p}))+O(\eta^4)\, .
\label{selfenergy}\end{eqnarray}
The vacuum bubble (diagram with no external lines, Fig.~\ref{feynman4}(c)) is
\begin{eqnarray}
V&=&2\mathrm{Re}\left\{\frac{1}{2\pi}\int\limits_{-\infty}^\infty dE \sum\limits_{\mathbf{p},\alpha}\frac{\Omega^2}{(c(\mathbf{k}_\mathrm{L}-\mathbf{p})+2\mathrm{i}E)(\mathrm{i}\nu_\alpha-\mathrm{i}E)}\left(\eta_\alpha \hat k_{\mathrm{L}\alpha}+\mathrm{i}\frac{d_\alpha(\mathbf{k}_\mathrm{L})-d_\alpha(\mathbf{k}_\mathrm{L}-\mathbf{p})}{c(\mathbf{k}_\mathrm{L})}\right)\right.\notag\\
&&\left.\times\left(-\frac{1}{2}\eta_\alpha \hat k_{\mathrm{L}\alpha}-\mathrm{i}\frac{d_\alpha(\mathbf{k}_\mathrm{L})-d_\alpha(\mathbf{k}_\mathrm{L}-\mathbf{p})}{2c(\mathbf{k}_\mathrm{L})}+\frac{\mathrm{Im}\,d_\alpha(\mathbf{k}_\mathrm{L})}{c^*(\mathbf{k}_\mathrm{L})}+\mathrm{i}\mathrm{Re}\,c(\mathbf{k}_\mathrm{L})\frac{d_\alpha(\mathbf{k}_\mathrm{L})-d_\alpha(\mathbf{k}_\mathrm{L}-\mathbf{p})}{|c(\mathbf{k}_\mathrm{L})|^2}\right)\right\}\notag\\
&&+2\mathrm{Re}\left\{\frac{\Omega^4}{\mathrm{i}\nu_\alpha|c(\mathbf{k}_\mathrm{L})|^4}\left(\mathrm{i}\eta_\alpha\hat k_{\mathrm{L}\alpha}\mathrm{Re}\,c(\mathbf{k}_\mathrm{L})-\mathrm{i}\mathrm{Im}\,d_\alpha(\mathbf{k}_\mathrm{L})\right)^2\right\}+O(\eta^4)\notag\\
&=&-\mathrm{Re}\sum\limits_{\mathbf{p},\alpha}M_{\mathrm{h} \alpha\alpha}(\mathbf{p})+O(\eta^4)\, ,\label{vacbubble}\end{eqnarray}
where the $1/2\pi$ in the loop energy integral is a Fourier transform normalization factor, and the $2\mathrm{Re}$ comes from including diagrams on both sides of the $\rho$ line.

Substituting these into Eq.~(\ref{fmaster}) and setting $\dot\rho=0$, we obtain Eq.~(\ref{cmp2}) at zeroth order and Eq.~(\ref{43}) at second order, i.e. the two methods agree.
\end{widetext}

\begin{thebibliography}{32}
\expandafter\ifx\csname natexlab\endcsname\relax\def\natexlab#1{#1}\fi
\expandafter\ifx\csname bibnamefont\endcsname\relax
  \def\bibnamefont#1{#1}\fi
\expandafter\ifx\csname bibfnamefont\endcsname\relax
  \def\bibfnamefont#1{#1}\fi
\expandafter\ifx\csname citenamefont\endcsname\relax
  \def\citenamefont#1{#1}\fi
\expandafter\ifx\csname url\endcsname\relax
  \def\url#1{\texttt{#1}}\fi
\expandafter\ifx\csname urlprefix\endcsname\relax\def\urlprefix{URL }\fi
\providecommand{\bibinfo}[2]{#2}
\providecommand{\eprint}[2][]{\url{#2}}

\bibitem[{\citenamefont{Wineland and Itano}(1979)}]{laser-cooling-limits}
\bibinfo{author}{\bibfnamefont{D.~J.} \bibnamefont{Wineland}} \bibnamefont{and}
  \bibinfo{author}{\bibfnamefont{W.~M.} \bibnamefont{Itano}},
  \bibinfo{journal}{Phys. Rev. A} \textbf{\bibinfo{volume}{20}},
  \bibinfo{pages}{1521} (\bibinfo{year}{1979}).

\bibitem[{\citenamefont{{Wallis} and
  {Ertmer}}(1989)}]{narrowline-doppler-theory}
\bibinfo{author}{\bibfnamefont{H.}~\bibnamefont{{Wallis}}} \bibnamefont{and}
  \bibinfo{author}{\bibfnamefont{W.}~\bibnamefont{{Ertmer}}},
  \bibinfo{journal}{J. Opt. Soc. Am. B} \textbf{\bibinfo{volume}{6}}, \bibinfo{pages}{2211}
  (\bibinfo{year}{1989}).

\bibitem[{\citenamefont{{Dalibard} and
  {Cohen-Tannoudji}}(1989)}]{polarization-gradient-theory}
\bibinfo{author}{\bibfnamefont{J.}~\bibnamefont{{Dalibard}}} \bibnamefont{and}
  \bibinfo{author}{\bibfnamefont{C.}~\bibnamefont{{Cohen-Tannoudji}}},
  \bibinfo{journal}{J. Opt. Soc. Am. B} \textbf{\bibinfo{volume}{6}}, \bibinfo{pages}{2023}
  (\bibinfo{year}{1989}).

\bibitem[{\citenamefont{Cirac et~al.}(1992)\citenamefont{Cirac, Blatt, Zoller,
  and Phillips}}]{sw-sideband-theory}
\bibinfo{author}{\bibfnamefont{J.~I.} \bibnamefont{Cirac}},
  \bibinfo{author}{\bibfnamefont{R.}~\bibnamefont{Blatt}},
  \bibinfo{author}{\bibfnamefont{P.}~\bibnamefont{Zoller}}, \bibnamefont{and}
  \bibinfo{author}{\bibfnamefont{W.~D.} \bibnamefont{Phillips}},
  \bibinfo{journal}{Phys. Rev. A} \textbf{\bibinfo{volume}{46}},
  \bibinfo{pages}{2668} (\bibinfo{year}{1992}).

\bibitem[{\citenamefont{Anderson et~al.}(1995)\citenamefont{Anderson, Ensher,
  Matthews, Wieman, and Cornell}}]{bec-orig-Rb}
\bibinfo{author}{\bibfnamefont{M.~H.} \bibnamefont{Anderson}},
  \bibinfo{author}{\bibfnamefont{J.~R.} \bibnamefont{Ensher}},
  \bibinfo{author}{\bibfnamefont{M.~R.} \bibnamefont{Matthews}},
  \bibinfo{author}{\bibfnamefont{C.~E.} \bibnamefont{Wieman}},
  \bibnamefont{and} \bibinfo{author}{\bibfnamefont{E.~A.}
  \bibnamefont{Cornell}}, \bibinfo{journal}{Science}
  \textbf{\bibinfo{volume}{269}}, \bibinfo{pages}{198} (\bibinfo{year}{1995}).

\bibitem[{\citenamefont{Davis et~al.}(1995)\citenamefont{Davis, Mewes, Andrews,
  van Druten, Durfee, Kurn, and Ketterle}}]{bec-orig-Na}
\bibinfo{author}{\bibfnamefont{K.~B.} \bibnamefont{Davis}},
  \bibinfo{author}{\bibfnamefont{M.~O.} \bibnamefont{Mewes}},
  \bibinfo{author}{\bibfnamefont{M.~R.} \bibnamefont{Andrews}},
  \bibinfo{author}{\bibfnamefont{N.~J.} \bibnamefont{van Druten}},
  \bibinfo{author}{\bibfnamefont{D.~S.} \bibnamefont{Durfee}},
  \bibinfo{author}{\bibfnamefont{D.~M.} \bibnamefont{Kurn}}, \bibnamefont{and}
  \bibinfo{author}{\bibfnamefont{W.}~\bibnamefont{Ketterle}},
  \bibinfo{journal}{Phys. Rev. Lett.} \textbf{\bibinfo{volume}{75}},
  \bibinfo{pages}{3969} (\bibinfo{year}{1995}).

\bibitem[{\citenamefont{Ido et~al.}(2000)\citenamefont{Ido, Isoya, and
  Katori}}]{narrowline-doppler-Sr}
\bibinfo{author}{\bibfnamefont{T.}~\bibnamefont{Ido}},
  \bibinfo{author}{\bibfnamefont{Y.}~\bibnamefont{Isoya}}, \bibnamefont{and}
  \bibinfo{author}{\bibfnamefont{H.}~\bibnamefont{Katori}},
  \bibinfo{journal}{Phys. Rev. A} \textbf{\bibinfo{volume}{61}},
  \bibinfo{pages}{061403(R)} (\bibinfo{year}{2000}).

\bibitem[{\citenamefont{Kasevich and Chu}(1992)}]{raman-doppler-Na}
\bibinfo{author}{\bibfnamefont{M.}~\bibnamefont{Kasevich}} \bibnamefont{and}
  \bibinfo{author}{\bibfnamefont{S.}~\bibnamefont{Chu}},
  \bibinfo{journal}{Phys. Rev. Lett.} \textbf{\bibinfo{volume}{69}},
  \bibinfo{pages}{1741} (\bibinfo{year}{1992}).

\bibitem[{\citenamefont{Reichel et~al.}(1995)\citenamefont{Reichel, Bardou,
  Ben Dahan, Peik, Rand, Salomon, and Cohen-Tannoudji}}]{raman-doppler-Cs}
\bibinfo{author}{\bibfnamefont{J.}~\bibnamefont{Reichel}},
  \bibinfo{author}{\bibfnamefont{F.}~\bibnamefont{Bardou}},
  \bibinfo{author}{\bibfnamefont{M.} \bibnamefont{Ben Dahan}},
  \bibinfo{author}{\bibfnamefont{E.}~\bibnamefont{Peik}},
  \bibinfo{author}{\bibfnamefont{S.}~\bibnamefont{Rand}},
  \bibinfo{author}{\bibfnamefont{C.}~\bibnamefont{Salomon}}, \bibnamefont{and}
  \bibinfo{author}{\bibfnamefont{C.}~\bibnamefont{Cohen-Tannoudji}},
  \bibinfo{journal}{Phys. Rev. Lett.} \textbf{\bibinfo{volume}{75}},
  \bibinfo{pages}{4575} (\bibinfo{year}{1995}).

\bibitem[{\citenamefont{Jaksch and Zoller}(2005)}]{lattice-review-Jaksch05}
\bibinfo{author}{\bibfnamefont{D.}~\bibnamefont{Jaksch}} \bibnamefont{and}
  \bibinfo{author}{\bibfnamefont{P.}~\bibnamefont{Zoller}},
  \bibinfo{journal}{Annals of Physics} \textbf{\bibinfo{volume}{315}},
  \bibinfo{pages}{52 } (\bibinfo{year}{2005}).

\bibitem[{\citenamefont{Yukalov}(2009)}]{lattice-review09}
\bibinfo{author}{\bibfnamefont{V.~I.} \bibnamefont{Yukalov}},
  \bibinfo{journal}{Laser Physics} \textbf{\bibinfo{volume}{19}},
  \bibinfo{pages}{1} (\bibinfo{year}{2009}), \eprint{arXiv:0901.0636}.

\bibitem[{\citenamefont{Lewenstein et~al.}(2007)\citenamefont{Lewenstein,
  Sanpera, Ahufinger, Damski, Sen(De), and Sen}}]{lattice-qs-review}
\bibinfo{author}{\bibfnamefont{M.}~\bibnamefont{Lewenstein}},
  \bibinfo{author}{\bibfnamefont{A.}~\bibnamefont{Sanpera}},
  \bibinfo{author}{\bibfnamefont{V.}~\bibnamefont{Ahufinger}},
  \bibinfo{author}{\bibfnamefont{B.}~\bibnamefont{Damski}},
  \bibinfo{author}{\bibfnamefont{A.}~\bibnamefont{Sen(De)}}, \bibnamefont{and}
  \bibinfo{author}{\bibfnamefont{U.}~\bibnamefont{Sen}},
  \bibinfo{journal}{Advances in Physics} \textbf{\bibinfo{volume}{56}},
  \bibinfo{pages}{243} (\bibinfo{year}{2007}).

\bibitem[{\citenamefont{Eschner et~al.}(2003)\citenamefont{Eschner, Morigi,
  Schmidt-Kaler, and Blatt}}]{ion-cooling-review}
\bibinfo{author}{\bibfnamefont{J.}~\bibnamefont{Eschner}},
  \bibinfo{author}{\bibfnamefont{G.}~\bibnamefont{Morigi}},
  \bibinfo{author}{\bibfnamefont{F.}~\bibnamefont{Schmidt-Kaler}},
  \bibnamefont{and} \bibinfo{author}{\bibfnamefont{R.}~\bibnamefont{Blatt}},
  \bibinfo{journal}{J. Opt. Soc. Am. B} \textbf{\bibinfo{volume}{20}}, \bibinfo{pages}{1003}
  (\bibinfo{year}{2003}).

\bibitem[{\citenamefont{Han et~al.}(2000)\citenamefont{Han, Wolf, Oliver,
  McCormick, DePue, and Weiss}}]{sideband-3dlattice-Cs}
\bibinfo{author}{\bibfnamefont{D.-J.} \bibnamefont{Han}},
  \bibinfo{author}{\bibfnamefont{S.}~\bibnamefont{Wolf}},
  \bibinfo{author}{\bibfnamefont{S.}~\bibnamefont{Oliver}},
  \bibinfo{author}{\bibfnamefont{C.}~\bibnamefont{McCormick}},
  \bibinfo{author}{\bibfnamefont{M.~T.} \bibnamefont{DePue}}, \bibnamefont{and}
  \bibinfo{author}{\bibfnamefont{D.~S.} \bibnamefont{Weiss}},
  \bibinfo{journal}{Phys. Rev. Lett.} \textbf{\bibinfo{volume}{85}},
  \bibinfo{pages}{724} (\bibinfo{year}{2000}).

\bibitem[{\citenamefont{Bouchoule et~al.}(2002)\citenamefont{Bouchoule,
  Morinaga, Salomon, and Petrov}}]{sideband-1dlattice-Cs}
\bibinfo{author}{\bibfnamefont{I.}~\bibnamefont{Bouchoule}},
  \bibinfo{author}{\bibfnamefont{M.}~\bibnamefont{Morinaga}},
  \bibinfo{author}{\bibfnamefont{C.}~\bibnamefont{Salomon}}, \bibnamefont{and}
  \bibinfo{author}{\bibfnamefont{D.~S.} \bibnamefont{Petrov}},
  \bibinfo{journal}{Phys. Rev. A} \textbf{\bibinfo{volume}{65}},
  \bibinfo{pages}{033402} (\bibinfo{year}{2002}).

\bibitem[{\citenamefont{Beige and Hegerfeldt}(1998)}]{two-dipole1}
\bibinfo{author}{\bibfnamefont{A.}~\bibnamefont{Beige}} \bibnamefont{and}
  \bibinfo{author}{\bibfnamefont{G.~C.} \bibnamefont{Hegerfeldt}},
  \bibinfo{journal}{Phys. Rev. A} \textbf{\bibinfo{volume}{58}},
  \bibinfo{pages}{4133} (\bibinfo{year}{1998}).

\bibitem[{\citenamefont{Sch\"on and Beige}(2001)}]{two-dipole2}
\bibinfo{author}{\bibfnamefont{C.}~\bibnamefont{Sch\"on}} \bibnamefont{and}
  \bibinfo{author}{\bibfnamefont{A.}~\bibnamefont{Beige}},
  \bibinfo{journal}{Phys. Rev. A} \textbf{\bibinfo{volume}{64}},
  \bibinfo{pages}{023806} (\bibinfo{year}{2001}).

\bibitem[{\citenamefont{Agarwal}(1974)}]{Agarwal}
\bibinfo{author}{\bibfnamefont{G.~S.} \bibnamefont{Agarwal}},
  \emph{\bibinfo{title}{Quantum Optics}} (\bibinfo{publisher}{Springer Tracts
  in Modern Physics vol. 70}, \bibinfo{year}{1974}).

\bibitem[{\citenamefont{Castin et~al.}(1998)\citenamefont{Castin, Cirac, and
  Lewenstein}}]{reabsorption-heating}
\bibinfo{author}{\bibfnamefont{Y.}~\bibnamefont{Castin}},
  \bibinfo{author}{\bibfnamefont{J.~I.} \bibnamefont{Cirac}}, \bibnamefont{and}
  \bibinfo{author}{\bibfnamefont{M.}~\bibnamefont{Lewenstein}},
  \bibinfo{journal}{Phys. Rev. Lett.} \textbf{\bibinfo{volume}{80}},
  \bibinfo{pages}{5305} (\bibinfo{year}{1998}).

\bibitem[{\citenamefont{Wunderlich et~al.}(2005)\citenamefont{Wunderlich,
  Morigi, and Rei\ss{}}}]{ion-chain-cooling}
\bibinfo{author}{\bibfnamefont{C.}~\bibnamefont{Wunderlich}},
  \bibinfo{author}{\bibfnamefont{G.}~\bibnamefont{Morigi}}, \bibnamefont{and}
  \bibinfo{author}{\bibfnamefont{D.}~\bibnamefont{Rei\ss{}}},
  \bibinfo{journal}{Phys. Rev. A} \textbf{\bibinfo{volume}{72}},
  \bibinfo{pages}{023421} (\bibinfo{year}{2005}).

\bibitem[{\citenamefont{Baranov}(2008)}]{dipole-gas-review08}
\bibinfo{author}{\bibfnamefont{M.}~\bibnamefont{Baranov}},
  \bibinfo{journal}{Physics Reports} \textbf{\bibinfo{volume}{464}},
  \bibinfo{pages}{71} (\bibinfo{year}{2008}).

\bibitem[{\citenamefont{Santos and
  Lewenstein}(1999)}]{collective-darkstate-cooling}
\bibinfo{author}{\bibfnamefont{L.}~\bibnamefont{Santos}} \bibnamefont{and}
  \bibinfo{author}{\bibfnamefont{M.}~\bibnamefont{Lewenstein}},
  \bibinfo{journal}{Phys. Rev. A} \textbf{\bibinfo{volume}{60}},
  \bibinfo{pages}{3851} (\bibinfo{year}{1999}).

\bibitem[{\citenamefont{Wolf et~al.}(2000)\citenamefont{Wolf, Oliver, and
  Weiss}}]{reabsorption-lattice}
\bibinfo{author}{\bibfnamefont{S.}~\bibnamefont{Wolf}},
  \bibinfo{author}{\bibfnamefont{S.~J.} \bibnamefont{Oliver}},
  \bibnamefont{and} \bibinfo{author}{\bibfnamefont{D.~S.} \bibnamefont{Weiss}},
  \bibinfo{journal}{Phys. Rev. Lett.} \textbf{\bibinfo{volume}{85}},
  \bibinfo{pages}{4249} (\bibinfo{year}{2000}).

\bibitem[{\citenamefont{Mendon\c{c}a et~al.}(2008)\citenamefont{Mendon\c{c}a,
  Kaiser, Ter\c{c}as, and Loureiro}}]{rescattering-force-theory}
\bibinfo{author}{\bibfnamefont{J.~T.} \bibnamefont{Mendon\c{c}a}},
  \bibinfo{author}{\bibfnamefont{R.}~\bibnamefont{Kaiser}},
  \bibinfo{author}{\bibfnamefont{H.}~\bibnamefont{Ter\c{c}as}},
  \bibnamefont{and} \bibinfo{author}{\bibfnamefont{J.}~\bibnamefont{Loureiro}},
  \bibinfo{journal}{Phys. Rev. A} \textbf{\bibinfo{volume}{78}},
  \bibinfo{eid}{013408} (\bibinfo{year}{2008}).

\bibitem[{\citenamefont{Hegerfeldt}(1993)}]{reset}
\bibinfo{author}{\bibfnamefont{G.~C.} \bibnamefont{Hegerfeldt}},
  \bibinfo{journal}{Phys. Rev. A} \textbf{\bibinfo{volume}{47}},
  \bibinfo{pages}{449} (\bibinfo{year}{1993}).

\bibitem[{\citenamefont{Holstein and Primakoff}(1940)}]{Holstein}
\bibinfo{author}{\bibfnamefont{T.}~\bibnamefont{Holstein}} \bibnamefont{and}
  \bibinfo{author}{\bibfnamefont{H.}~\bibnamefont{Primakoff}},
  \bibinfo{journal}{Phys. Rev.} \textbf{\bibinfo{volume}{58}},
  \bibinfo{pages}{1098} (\bibinfo{year}{1940}).

\bibitem[{\citenamefont{Shah et~al.}(1974)\citenamefont{Shah, Umezawa, and
  Vitiello}}]{Shah}
\bibinfo{author}{\bibfnamefont{M.~N.} \bibnamefont{Shah}},
  \bibinfo{author}{\bibfnamefont{H.}~\bibnamefont{Umezawa}}, \bibnamefont{and}
  \bibinfo{author}{\bibfnamefont{G.}~\bibnamefont{Vitiello}},
  \bibinfo{journal}{Phys. Rev. B} \textbf{\bibinfo{volume}{10}},
  \bibinfo{pages}{4724} (\bibinfo{year}{1974}).

\bibitem[{\citenamefont{Antezza and Castin}(2009)}]{polariton-3dlattice}
\bibinfo{author}{\bibfnamefont{M.}~\bibnamefont{Antezza}} \bibnamefont{and}
  \bibinfo{author}{\bibfnamefont{Y.}~\bibnamefont{Castin}},
  \bibinfo{journal}{Phys. Rev. Lett.} \textbf{\bibinfo{volume}{103}},
  \bibinfo{pages}{123903} (\bibinfo{year}{2009}).

\bibitem[{\citenamefont{Keldysh}(1964)}]{keldysh-orig}
\bibinfo{author}{\bibfnamefont{L.~V.} \bibnamefont{Keldysh}},
  \bibinfo{journal}{Zh. Eksp. Teor. Fiz.} \textbf{\bibinfo{volume}{47}},
  \bibinfo{pages}{1515} (\bibinfo{year}{1964}).

\bibitem[{\citenamefont{Rammer and Smith}(1986)}]{keldysh-review}
\bibinfo{author}{\bibfnamefont{J.}~\bibnamefont{Rammer}} \bibnamefont{and}
  \bibinfo{author}{\bibfnamefont{H.}~\bibnamefont{Smith}},
  \bibinfo{journal}{Rev. Mod. Phys.} \textbf{\bibinfo{volume}{58}},
  \bibinfo{pages}{323} (\bibinfo{year}{1986}).

\bibitem[{\citenamefont{Dyson}(1949)}]{dyson1}
\bibinfo{author}{\bibfnamefont{F.~J.} \bibnamefont{Dyson}},
  \bibinfo{journal}{Phys. Rev.} \textbf{\bibinfo{volume}{75}},
  \bibinfo{pages}{486} (\bibinfo{year}{1949}).

\bibitem[{\citenamefont{Stenholm}(1986)}]{laser-cooling-theory}
\bibinfo{author}{\bibfnamefont{S.}~\bibnamefont{Stenholm}},
  \bibinfo{journal}{Rev. Mod. Phys.} \textbf{\bibinfo{volume}{58}},
  \bibinfo{pages}{699} (\bibinfo{year}{1986}).

\end{thebibliography}
\end{document}